\documentclass[journal]{IEEEtran}

\usepackage{amsmath,amsfonts}

\usepackage{algorithm}
\usepackage{algpseudocode}

\usepackage{array}                    % ✅ 1. First
\usepackage[table]{xcolor}            % ✅ 2. Then xcolor
\definecolor{lightgray}{gray}{0.9}    % ✅ 3. Define colors
\usepackage{colortbl}                 % ✅ 4. Then colortbl
\usepackage{booktabs}                 % ✅ 5. Then booktabs
\usepackage{tabularx}                 % ✅ 6. Then tabularx

\usepackage{graphicx}
\usepackage[caption=false,font=normalsize,labelfont=sf,textfont=sf]{subfig}
\usepackage{caption}
\usepackage{float}

\usepackage{textcomp}
\usepackage{url}
\usepackage{verbatim}
\usepackage{enumitem}

\usepackage{cite}

\usepackage{placeins}

\begin{document}
\title{Multi-agent deep reinforcement learning for UAV-based 5G network slicing}
\author{
\IEEEauthorblockN{Ghoshana Bista\IEEEauthorrefmark{1}, 
Abbas Bradai\IEEEauthorrefmark{2}, 
Emmanuel Moulay\IEEEauthorrefmark{1}, 
Abdulhalim Dandoush\IEEEauthorrefmark{3}}

\IEEEauthorblockA{\IEEEauthorrefmark{1}XLIM Institute (UMR CNRS 7252), Université de Poitiers, France\\
Email: \{ghoshana.bista, emmanuel.moulay\}@univ-poitiers.fr}

\IEEEauthorblockA{\IEEEauthorrefmark{2}LEAT Lab (UMR CNRS 7248), Université Côte d'Azur, France\\
Email: abbas.bradai@univ-cotedazur.fr}

\IEEEauthorblockA{\IEEEauthorrefmark{3}University of Doha for Science and Technology, Qatar\\
Email: abdulhalim.dandoush@udst.edu.qa}
}

\maketitle

%\author{IEEE Publication Technology,~\IEEEmembership{Staff,~IEEE,}
        % <-this % stops a space
%\thanks{This paper was produced by the IEEE Publication Technology Group. They are in Piscataway, NJ.}% <-this % stops a space
%\thanks{Manuscript received April 19, 2021; revised August 16, 2021.}
%}

%\markboth{Journal of GB}%Journal of \LaTeX\ Class Files,~Vol.~14, No.~8, August~2021}%
%{Shell \MakeLowercase{\textit{et al.}}:"Dynamic Multi-Agent Reinforcement Learning for UAV-based Communication: Comparative Analysis of PPO, DDPG, and DQN in Urban and Rural Scenarios"}


%\maketitle

\begin{abstract}
The growing demand for robust and scalable wireless networks in the 5G-and-beyond era has led to the deployment of Unmanned Aerial Vehicles (UAVs) as mobile base stations to enhance coverage in dense urban and underserved rural environments. This paper presents a Multi-Agent Deep Reinforcement Learning (MADRL) framework that integrates Proximal Policy Optimization (MAPPO), Multi-Agent Deep Deterministic Policy Gradient (MADDPG), and Multi-Agent Deep Q-Networks (MADQN) to jointly optimize UAV positioning, resource allocation, Quality of Service (QoS), and energy efficiency through 5G network slicing. The framework adopts Centralized Training with Decentralized Execution (CTDE), enabling autonomous real-time decision-making while preserving global coordination.

Users are prioritized into Premium $(A)$, Silver $(B)$, and Bronze $(C)$ slices with distinct QoS requirements. Experiments in realistic urban and rural scenarios show that MAPPO achieves the best overall QoS–energy tradeoff, especially in interference-rich environments; MADDPG offers more precise continuous control and can attain slightly higher SINR in open rural settings at the cost of increased energy usage; and MADQN provides a computationally efficient baseline for discretized action spaces. These findings demonstrate that no single MARL algorithm is universally dominant; instead, algorithm suitability depends on environmental topology, user density, and service requirements. The proposed framework highlights the potential of MARL-driven UAV systems to enhance scalability, reliability, and differentiated QoS delivery in next-generation wireless networks.
\end{abstract}

%The increasing demand for robust and scalable wireless networks in the 5G era has driven the deployment of Unmanned Aerial Vehicles (UAVs) as mobile base stations to address coverage gaps in dense urban and underserved rural environments. This paper presents a novel Multi-Agent Deep Reinforcement Learning (MADRL) framework that integrates Deep Deterministic Policy Gradient (DDPG), Proximal Policy Optimization (PPO), and Deep Q-Network (DQN) algorithms to optimize UAV positioning and resource allocation for heterogeneous Quality of Service (QoS) delivery, and energy consumption,  leveraging 5G network slicing. The framework uses Centralized Training with Decentralized Execution (CTDE) to enable real-time, autonomous UAV decision-making while ensuring global coordination.
%We evaluate the MADRL framework in urban and rural scenarios, incorporating user prioritization into 5G network slicing, where users are categorized into Premium (A), Silver (B), and Bronze (C) classes. Results show thatMADDPG excels in high QoS precision and energy efficiency, MAPPO is optimal for interference-rich environments, andMADQN demonstrates computational efficiency in stable, low-variability settings. These findings highlight the potential of MARL-driven UAV systems to enhance wireless network scalability, reliability, and QoS delivery in diverse real-world applications.


%%%%%%%%%%%%%%%%%%%%%%%%%%%%%%%%%%%%%%%%%%%%%%%%%%%%%%%%%%
\section{Introduction}
%%%%%%%%%%%%%%%%%%%%%%%%%%%%%%%%%%%%%%%%%%%%%%%%%%%%%%%%%%
\IEEEPARstart{T}{he} rapid expansion of wireless communication networks, driven by the proliferation of Internet of Things (IoT) devices and 5G technology, has intensified the demand for flexible and adaptive wireless coverage. Traditional ground-based base stations often struggle in highly dynamic environments: urban regions suffer from dense user populations, severe interference, and frequent coverage holes, while rural areas lack sufficient infrastructure to guarantee reliable connectivity. These limitations underscore the need for agile and scalable solutions capable of delivering differentiated service across heterogeneous environments.

Unmanned Aerial Vehicles (UAVs) have emerged as a promising complement to terrestrial infrastructure. Acting as mobile base stations, UAVs offer rapid deployment, enhanced maneuverability, and the ability to dynamically reposition to meet short-term coverage demands or disaster-induced outages~\cite{zeng2016wireless,shi2018drone,savkin2020navigation}. However, ensuring high Quality of Service (QoS) with UAVs requires solving tightly coupled challenges related to mobility, interference, limited battery capacity, and fluctuating traffic loads. These challenges become even more prominent in future 5G-and-beyond architectures, where multi-service differentiation and stringent QoS guarantees must be provided simultaneously.

To address these requirements, 5G network slicing allows physical resources to be partitioned into virtual slices with distinct QoS guarantees. In UAV-assisted networks, this capability is essential for serving heterogeneous users—Premium ($A$), Silver ($B$), and Bronze ($C$)—each with unique latency, throughput, and SINR targets. Efficient UAV deployment therefore requires the joint optimization of slice-level QoS, UAV motion, energy use, and interference management~\cite{shamsoshoara2024joint}.

Traditional optimization and control approaches struggle to meet these combined objectives due to the highly non-stationary and multi-agent nature of UAV networks. Reinforcement Learning (RL), and particularly Multi-Agent Reinforcement Learning (MARL), offers a compelling alternative due to its ability to learn adaptive policies in complex environments. While prior MARL studies have investigated UAV coverage, trajectory planning, or throughput optimization, they typically address a single metric and do not incorporate slice-based QoS differentiation or joint QoS–energy optimization.

\textbf{Contributions.}  
In this work, we propose a unified Multi-Agent Deep Reinforcement Learning (MADRL) framework that integrates 5G network slicing with coordinated UAV control under a Centralized Training with Decentralized Execution (CTDE) paradigm. Our framework jointly optimizes latency, throughput, SINR, and energy consumption across user priority levels ($A/B/C$), addressing a multi-slice QoS problem that remains underexplored in UAV-assisted networks. Specifically:
\begin{itemize}
    \item We develop a CTDE-based MADRL architecture enabling UAVs to autonomously adapt movement, coverage, and resource allocation while respecting slice-specific QoS requirements.
    \item We integrate three representative MADRL algorithms—MAPPO, MADDPG, and MADQN—chosen for their ability to handle continuous, mixed, and discrete action spaces, respectively.
    \item We provide a comparative analysis of these algorithms in two contrasting environments (dense urban and sparse rural), highlighting how environmental structure affects MARL suitability rather than assuming a single universal solution.
\end{itemize}

\textbf{Findings.}  
Our experiments reveal a clear scenario-dependent pattern. MAPPO achieves the best overall QoS--energy tradeoff, especially in interference-rich urban environments requiring coordinated continuous control. MADDPG provides smoother trajectories and can yield slightly higher SINR in open rural areas, but at the cost of higher energy consumption and variance. MADQN offers computational efficiency but underperforms in complex or highly dynamic environments due to its discretized action space. These results emphasize that algorithm selection must consider both environmental topology and QoS requirements rather than relying on a single MARL method.

\textbf{Paper organization.}  
Section~\ref{Sec:bib} reviews related work. Section~\ref{Sec:mod} presents the system modeling and channel assumptions. Section~\ref{Sec:optim} formulates the UAV QoS--energy optimization problem. Section~\ref{Sec:MADRL} details the MARL algorithms and training approach. Section~\ref{Sec:results} evaluates their performance in urban and rural scenarios. Section~\ref{Sec:conclusion} concludes the paper.
For implementation details, please refer to our code repository.\footnote{%
\url{https://github.com/ghoshana/MARL_COMPARE}.
If you cannot access it, please contact the first author and I will provide access.}
%%%%%%%%%%%%%%%%%%%%%%%%%%%%%%%%%%%%%%%%%%%%%%%%%%%%%%%%%%
\section{Bibliographical review}\label{Sec:bib}
%%%%%%%%%%%%%%%%%%%%%%%%%%%%%%%%%%%%%%%%%%%%%%%%%%%%%%%%%%
UAVs are increasingly integral to modern communication networks, addressing scalability demands. DRL enables UAVs to autonomously optimize tasks such as path planning and resource allocation, significantly enhancing QoS \cite{azar2021drone}. Recent advances in DRL have demonstrated notable improvements in UAV trajectory planning and resource management, with research broadly categorized into UAV operations and QoS optimization.
%%%%%%%%%%%%%%%%%%%%%%%%%%%%%%%%%%%%%%%%%%%%%%%%%%%%%%%%%%
\subsection{DRL in UAV operations.}
The application of DRL in UAV operations has shifted network infrastructure management paradigms. For example, Shamsoshoara et al. propose in \cite{Shamsoshoara2023JointPP} an interference-aware path planning and power allocation mechanism for UAVs using inverse reinforcement learning. The approach, compared against Behavioral Cloning and other methods, demonstrates improved performance, though scalability in more complex networks requires further investigation. Jiang et al. employ in \cite{jiang2021marl} a MARL framework to optimize UAV base station positions within Vehicular Ad-Hoc Networks (VANETs). While the framework reduces data transmission delays, challenges remain regarding real-time scalability in complex environments. Similarly, Kim et al. present in \cite{kim2024cooperative} a multi-agent DRL framework for enhancing energy management and network efficiency, but scalability in larger networks with more UAVs and users requires further study.
DRL applications in Internet of Drones (IoD) networks tackle challenges such as navigation and channel allocation, but practical concerns like real-time decision-making and computational complexity are underexplored \cite{article}. Tarekegn et al. demonstrate in \cite{tarekegn2022deep} how DRL can optimize drone base station deployment for improved coverage, although multi-drone scenarios and energy demands remain areas for future research. Hammami et al. investigate in \cite{hammami2019drone} MARL for managing drone-assisted cellular networks, focusing on deployment during peak demand periods, but larger network scalability remains unresolved. Warrier et al. use in \cite{warrier2022interference} a DQL framework to mitigate air-ground interference in 5G-connected UAVs, though computational and latency challenges persist. Additionally, Ouamri et al. propose in \cite{ouamri2023joint} a DDPG-based solution to optimize energy efficiency and throughput in UAV-assisted wireless power transfer. However, the method’s scalability in complex networks is not fully addressed.
Similarly, Zhou et al. present in \cite{zhou2021qoe} a hybrid DRL strategy for UAV deployment to improve Quality of Experience (QoE), but the computational demands of the hybrid model limit real-time application. Baghdady et al. explore in \cite{baghdady2024reinforcement} 3D positioning and frequency allocation for UAV coverage optimization but do not sufficiently address scalability for larger networks. In \cite{tariq2024towards}, Tariq et al. aim to reduce computational costs through task offloading from mobile users to UAVs, but the algorithm’s real-world scalability remains an open question. Pogaku et al. investigate in \cite{pogaku2022uav} UAV-assisted reconfigurable intelligent surfaces (RIS) to enhance energy efficiency and signal quality. However, integration challenges in urban environments with high mobility need further exploration. Cheng et al. \cite{cheng2024high} proposed a high-sample-efficient MARL framework to enhance adaptability in dynamic multitask UAV swarm environments. This work supports our focus on scalability and generalizability for large-scale UAV operations.

%%%%%%%%%%%%%%%%%%%%%%%%%%%%%%%%%%%%%%%%%%%%%%%%%%%%%%%%%%
\subsection{UAVs and QoS in communication networks.}
Integrating UAVs into communication networks has significantly improved QoS, particularly for 6G systems. Ni et al. introduce in \cite{ni2024throughput} a fairness control scheme for UAV-enabled wireless communication, optimizing resource allocation and UAV trajectory for balanced throughput and fairness in 6G networks. However, the study must sufficiently explore scalability in dynamic or large-scale networks. Tripathi et al. utilize in \cite{tripathi2022socially} social relationships among IoT devices for optimal UAV trajectory planning and 3D beamforming, though scalability in larger networks still needs to be fully addressed. Zhang et al. propose in \cite{zhang2021aoi} an Age of Information (AoI)-driven framework for UAV-assisted 6G networks, optimizing delay and error rates, but the challenges of implementing finite blocklength coding in dynamic environments require further investigation. In \cite{nasr2023distillation}, Nasr et al. leverage federated learning and distillation techniques to optimize UAV trajectories, improving QoS while ensuring data security, though scalability in complex networks remains a concern. In \cite{bose2022improving}, Bose et al. focus on reducing delays and energy consumption but still need to fully address scalability in large networks. Wang et al. propose in \cite{wang2020intelligent} a Mobile Edge Computing (MEC) solution for UAVs to enhance network performance and extend sensor network lifetimes, though larger-scale application remains unexplored. Finally, Gil et al. develop in \cite{gil2023optimizing} a mixed-integer linear programming model for IoT microservice placement on UAVs aimed at reducing latency. However, the solution's scalability for more complex networks needs to be examined.
The use of MADRL for trajectory optimization in UAV networks has also been explored by Zhang et al., who employed UAV jammers for secure multi-UAV communication \cite{zhang2020uav}. While their focus is on security, our approach targets QoS management in 5G networks, showcasing the broad applicability and flexibility of MADRL in UAV scenarios. 
In summary, while UAV-assisted networks and DRL-based optimization techniques have shown promise, key challenges persist in scaling these solutions for large, complex networks. Our research addresses these gaps by proposing a novel DRL-driven framework that prioritizes users by QoS requirements, contributing to more scalable UAV-based 5G network slicing.

%Geng et al. \cite{geng2024joint} explored joint UAV deployment and resource allocation using MINLP, which aligns with our method of optimizing UAV positioning to enhance QoS in dynamic environments.

In Table~\ref{tab:abbreviations}, we provide the abbreviations used in this article.
\begin{table}[htbp]
\centering
\caption{List of Abbreviations}
\begin{tabular}{|l|l|}
\hline
\rowcolor{lightgray} \textbf{Abbreviations} & \textbf{Definitions} \\ \hline
COMA                 & Counterfactual Multi-Agent Policy Gradient \\ \hline
CTDE                 & Centralized Training with Decentralized Execution \\ \hline
D2BS                 & Drone-to-Base Station \\ \hline
D2D                  & Drone-to-Drone \\ \hline
DDPG                 & Deep Deterministic Policy Gradient \\ \hline
DQN                  & Deep Q-Network \\ \hline
DRL                  & Deep Reinforcement Learning \\ \hline
LoS                  & Line of Sight \\ \hline
MADDPG               & Multi-Agent Deep Deterministic Policy Gradient \\ \hline
MADRL                & Multi-Agent Deep Reinforcement Learning \\ \hline
NLoS                 & Non-Line of Sight \\ \hline
PPO                  & Proximal Policy Optimization \\ \hline
QMIX                 & QMIX Algorithm (a value-based MARL method) \\ \hline
QoS                  & Quality of Service \\ \hline
RL                   & Reinforcement Learning \\ \hline
SINR                 & Signal-to-Interference-plus-Noise Ratio \\ \hline
UAV                  & Unmanned Aerial Vehicle \\ \hline
\end{tabular}
\label{tab:abbreviations}
\end{table}
%%%%%%%%%%%%%%%%%%%%%%%%%%%%%%%%%%%%%%%%%%%%%%%%%%%%%%%%%%
\section{System modeling}\label{Sec:mod}
\begin{figure}[!htbp]
    \centering
    \includegraphics[width=0.48\textwidth]{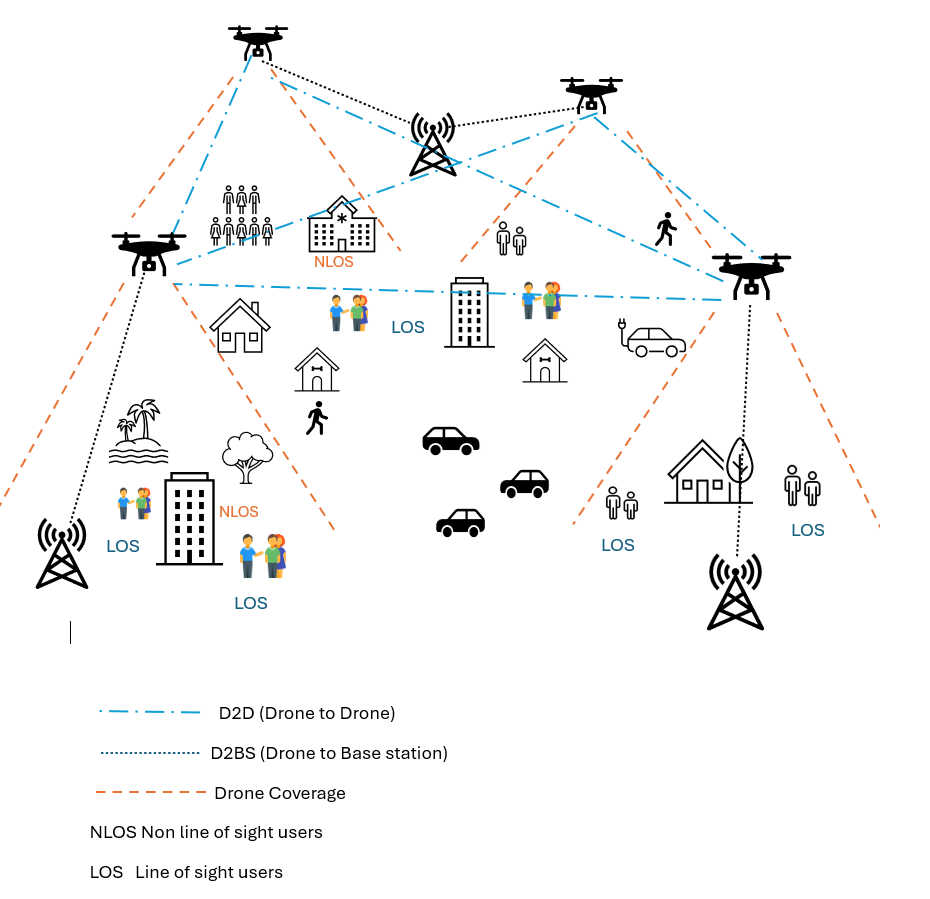}
    \caption{UAV-based 5G network}
    \label{fig:enter-label}
\end{figure}

%%%%%%%%%%%%%%%%%%%%%%%%%%%%%%%%%%%%%%%%%%%%%%%%%%%%%%%%%%
We consider a multi-drone communication system, as depicted in Figure~\ref{fig:enter-label}, integrating Drone-to-Drone (D2D) and Drone-to-Base Station (D2BS) links under Line-of-Sight (LoS) and Non-Line-of-Sight (NLoS) conditions. Each drone acts as an aerial base station, providing coverage to ground users and exchanging data with neighboring drones and the fixed infrastructure. User locations are assumed to be known via GPS, while drones move in 3D and users move in 2D. A centralized training with decentralized execution (CTDE) framework is employed to optimize throughput, SINR, and latency, while minimizing UAV energy consumption and enforcing user-priority constraints.

%%%%%%%%%%%%%%%%%%%%%%%%%%%%%%%%%%%%%%%%%%%%%%%%%%%%%%%%%%
\subsubsection{5G network slicing}
The system utilizes 5G network slicing to differentiate resource allocation among user types in both urban and rural environments. Drones are deployed in an $N_g \times N_g$ grid to serve Premium ($A$), Silver ($B$), and Bronze ($C$) users. Slice $A$ is given the highest priority, while slices $B$ and $C$ are allocated resources so as to maintain acceptable QoS under resource constraints. This enables differentiated service and efficient resource allocation, adapting to the specific demands of dense urban areas and wide rural coverage regions~\cite{messaoud2020deep}.

%%%%%%%%%%%%%%%%%%%%%%%%%%%%%%%%%%%%%%%%%%%%%%%%%%%%%%%%%%
\subsection{UAV communications}
%%%%%%%%%%%%%%%%%%%%%%%%%%%%%%%%%%%%%%%%%%%%%%%%%%%%%%%%%%

\subsubsection{Channel model}
All wireless links are modeled using a simplified air-to-ground channel inspired by the 3GPP TR~38.901 framework. For each UAV–user pair, we distinguish between LoS and NLoS conditions and compute large-scale path loss with additional shadowing and small-scale fading.

In practice, we use a 3GPP-TR38.901-based channel implementation, which, for each link, (i) draws a LoS/NLoS state according to a distance-dependent probability, (ii) applies the corresponding LoS/NLoS path-loss expression, and (iii) adds log-normal shadowing and Rician fading. This provides a realistic approximation of urban and rural air-to-ground propagation conditions and is consistent with our previous work~\cite{zhang2019framework,li2018deployment,albrecht2024multi,luong2021deep,bista2024differentiated}.

\subsubsection{Drone-to-Drone (D2D) communication}
Drone-to-drone (D2D) communication is mainly used for coordination and information exchange (e.g., user statistics, local QoS indicators). These links are predominantly LoS and are modeled using the same large-scale channel as for UAV–user links, but with the distance between UAVs. Additional shadowing is included to capture residual environmental effects, while small-scale fading is modeled consistently with the air-to-ground channel~\cite{zhang2019framework,li2018deployment}.

\subsubsection{Drone-to-Base Station (D2BS) communication}
For D2BS communication, we assume the presence of a reliable backhaul link as in~\cite{zhang2019framework,li2018deployment}. The propagation model again distinguishes LoS and NLoS components for rural and urban conditions. Rural areas exhibit a higher LoS probability, leading to stronger direct components and lower shadowing variance. In contrast, urban areas with numerous obstacles are characterized by a lower effective Rician $K$-factor and higher shadowing, capturing stronger multipath and attenuation effects. In the simulations, these effects are absorbed into the same TR38.901-inspired channel implementation used for D2U and D2D links.

%%%%%%%%%%%%%%%%%%%%%%%%%%%%%%%%%%%%%%%%%%%%%%%%%%%%%%%%%%
\subsection{Simulation setup and design} %Simulation environment design
%%%%%%%%%%%%%%%%%%%%%%%%%%%%%%%%%%%%%%%%%%%%%%%%%%%%%%%%%%
This study investigates UAV-based 5G network slicing in two deployment scenarios: an interference-rich urban environment and a sparsely populated rural area. In both cases, Premium (type~$A$), Silver (type~$B$), and Bronze (type~$C$) users coexist and request service from the UAVs. Urban areas are characterized by dense obstacles and high user density, whereas rural areas have fewer obstacles and more dispersed users. The propagation environment in both scenarios is modeled using a TR~38.901-inspired channel, incorporating distance-dependent path loss, log-normal shadowing, and Rician fading.

\paragraph{User mobility and coverage}
Users follow a random-walk or Markovian mobility model with random directions and speeds. In urban areas, higher population density leads to more frequent user movement and cluster formation, while in rural areas users are spread over a larger region and often move less (e.g., remaining at home or in fields). These distinct behaviors are implemented via separate urban and rural grid models.

Drone movement is modeled as
\begin{equation}\label{constraint_DM}
s_i(t+1) = s_i(t) + a_i(t), \quad \forall i \in \{1, \dots, N\},
\end{equation}
where $s_i(t)=(s_i^x(t),s_i^y(t),s_i^z(t))\in\mathbb{R}^3_+$ is the position of drone $i$ at time $t$, and $a_i(t)\in\mathbb{R}^3_+$ is the action selected by the MARL policy.

\paragraph{Drone deployment and communication}
Drones are initially positioned on a 3D grid and then controlled by the MARL policy to provide wireless coverage. In the urban scenario, more frequent repositioning is required to overcome NLoS conditions and coverage holes caused by obstacles; in the rural scenario, extended communication ranges allow fewer drones to cover a larger area. A mobile charging station is included and periodically relocated toward the centroid of active user clusters, reducing travel distance for low-battery drones and improving energy efficiency.

\paragraph{Key metrics and user prioritization}
Key QoS metrics evaluated in both scenarios include latency, throughput, and SINR (Signal-to-Interference-plus-Noise Ratio). The UAV network prioritizes Premium users (type~$A$) to ensure stringent QoS, such as low latency and high throughput, while providing relaxed QoS targets for types~$B$ and $C$. Drones assess local user density and QoS levels, adapt their movement and association decisions, and use collision-avoidance strategies to maintain safe inter-drone distances. Interference and SINR are derived from the TR~38.901-inspired channel model, as detailed in~\cite{bista2024differentiated}.

%%%%%%%%%%%%%%%%%%%%%%%%%%%%%%%%%%%%%%%%%%%%%%%%%%%%%%%%%%
\section{Optimization problem}\label{Sec:optim}
%%%%%%%%%%%%%%%%%%%%%%%%%%%%%%%%%%%%%%%%%%%%%%%%%%%%%%%%%%
In UAV-based wireless communication systems, optimizing QoS for users while reducing drone energy consumption is a critical challenge. This section formulates the multi-objective optimization problem for a fleet of $N$ drones deployed to serve users of priority types $A$, $B$, and $C$ while minimizing drone energy usage. The key difficulty is to balance resource allocation, drone movement, and energy efficiency, ensuring continuous coverage and fairness among all user types.

%%%%%%%%%%%%%%%%%%%%%%%%%%%%%%%%%%%%%%%%%%%%%%%%%%%%%%%%%%
\subsection{QoS and energy consumption}
%%%%%%%%%%%%%%%%%%%%%%%%%%%%%%%%%%%%%%%%%%%%%%%%%%%%%%%%%%
The proposed framework aims to maximize overall network utility while minimizing energy consumption. The instantaneous utility of drone $i \in \{1,\ldots,N\}$ is defined as a weighted sum of per-type QoS scores:
\begin{equation}
U_i(t) = \omega_A Q_i^A(t) + \omega_B Q_i^B(t) + \omega_C Q_i^C(t),
\end{equation}
where $\omega_A$, $\omega_B$, and $\omega_C$ are priority weights for user types $A$, $B$, and $C$, respectively, with $\omega_A > \omega_B > \omega_C$. The quantities $Q_i^A(t)$, $Q_i^B(t)$, and $Q_i^C(t)$ denote the QoS scores associated with the users of each type that are served by drone $i$. The system prioritizes high-priority users (type~$A$), aiming to minimize their latency and enhance their available bandwidth, even under energy constraints.

For a user type $X \in \{A,B,C\}$ served by drone $i$, the normalized QoS score $Q_i^X(t)$ is defined as
\begin{align}
    Q_i^X(t) = 
      \frac{L_{\text{target}}^X - L_i^X(t)}{L_{\text{target}}^X}
    + \frac{T_i^X(t) - T_{\text{target}}^X}{T_{\text{target}}^X}
    + \frac{S_i^X(t) - S_{\text{target}}^X}{S_{\text{target}}^X},
    \label{equation:qos_metric_updated}
\end{align}
where $L_i^X(t)$, $T_i^X(t)$, and $S_i^X(t)$ are, respectively, the latency, throughput, and SINR experienced by user type $X$ when served by drone $i$ at time $t$, and $L_{\text{target}}^X$, $T_{\text{target}}^X$, and $S_{\text{target}}^X$ are the corresponding target values. Each QoS component is normalized around its type-dependent target and summed into a single scalar score.

\paragraph{Energy model abstraction}
The energy model tracks a normalized battery level with a hover cost, a
velocity-proportional term, and a per-user service cost. This surrogate model
is designed to expose QoS–energy trade-offs rather than to emulate a specific
UAV platform. The ``moving charging station'' represents an abstract
service facility (e.g., a relocatable ground charger), without explicit
relocation cost modeling. These abstractions keep the environment lightweight
while preserving the algorithmic comparison objective of the study.

%%%%%%%%%%%%%%%%%%%%%%%%%%%%%%%%%%%%%%%%%%%%%%
\subsection{Drone operational constraints}
%%%%%%%%%%%%%%%%%%%%%%%%%%%%%%%%%%%%%%%%%%%%%%
Drones collaborate to optimize coverage, reduce redundancy, and conserve energy by sharing positional and user data while minimizing overlap. Each drone operates under several constraints:

\paragraph{Battery capacity}
Drones have limited battery capacity and must return to charging stations if their energy falls below a threshold $B_{\min}$:
\begin{equation}\label{constraint_BC}
 B_i(t) \geq B_{\min}, \quad \forall i \in \{1, \dots, N\},
\end{equation}
where $B_i(t)$ is the battery level of drone $i$ at time $t$.

\paragraph{Coverage}
Drones must maintain coverage for users within their communication range $d_{\text{com}}$:
\begin{equation}\label{constraint_CR}
 \| s_i(t) - u_m(t) \| \leq d_{\text{com}}, \quad \forall m \in \{1, \dots, M_i\},
\end{equation}
where $s_i(t)\in\mathbb{R}^3_+$ is the position of drone $i$ in $(x,y,z)$-coordinates, $u_m(t)\in\mathbb{R}^3_+$ is the position of user $m$ at time $t$, and $M_i$ is the number of users served by drone $i$. The total number of users is $M=\sum_{i=1}^N M_i$.

\paragraph{Altitude limits}
Drones must operate within predefined altitude boundaries:
\begin{equation}\label{constraint_H}
    h_i^{\min} \leq s_i^z(t) \leq h_i^{\max}, \quad \forall i \in \{1, \dots, N\},
\end{equation}
where $s_i^z(t)$ is the altitude of drone $i$, and $h_i^{\min}$ and $h_i^{\max}$ are the minimum and maximum allowable altitudes.

\paragraph{Collision avoidance}
Drones must maintain a safe distance $d_{\min}$ from one another to prevent collisions:
\begin{equation}\label{constraint_CA}
    \| s_i(t) - s_j(t) \| \geq d_{\min}, \quad \forall i, j \in \{1, \dots, N\}, \; i \neq j,
\end{equation}
where $d_{\min}$ is the minimum safety distance between drones.

\paragraph{Collaboration}
To avoid redundancy, drones must ensure efficient user coverage without serving the same users simultaneously:
\begin{equation}\label{constraint_CoL}
    M_i(t) \cap M_j(t) = \emptyset, \quad \forall i, j \in \{1, \dots, N\}, \; i \neq j,
\end{equation}
where $M_i(t)$ and $M_j(t)$ are the sets of users served by drones $i$ and $j$ at time $t$.

\paragraph{Drone movement}
Drones must adapt smoothly to changes in user distribution:
\begin{equation}\label{constraint_DV}
    \| s_i(t+1) - s_i(t) \| \leq v_{\max} \cdot \Delta t, \quad \forall i \in \{1, \dots, N\},
\end{equation}
where $v_{\max}$ is the maximum allowable velocity and $\Delta t$ is the time step.

\paragraph{Charging station adjustment}
Charging station positions dynamically adjust based on user density:
\begin{equation}\label{constraint_DCS}
    \| c(t+1) - c(t) \| \leq \delta_{\text{station}}, \quad \forall t,
\end{equation}
where $c(t)$ is the charging station position at time $t$, and $\delta_{\text{station}}$ is the maximum adjustment per step.

%%%%%%%%%%%%%%%%%%%%%%%%%%%%%%%%%%%%%%%%%%%%%%%%%%%%%%%%%%
\subsection{Mathematical formulation}
%%%%%%%%%%%%%%%%%%%%%%%%%%%%%%%%%%%%%%%%%%%%%%%%%%%%%%%%%%
Combining QoS and energy considerations, the multi-objective problem is written in scalarized form as
\begin{align}
    \max_{a(t)} \; & 
    \sum_{i=1}^{N}\Big( \omega_A Q_i^A(t) + \omega_B Q_i^B(t) + \omega_C Q_i^C(t) \Big)
    - \lambda \sum_{i=1}^{N} E_i(t) \label{equation:optimisation}\\
    \text{s.t.}\; &
    \eqref{constraint_BC},\;
    \eqref{constraint_CR},\;
    \eqref{constraint_H},\;
    \eqref{constraint_DM},\;
    \eqref{constraint_DV},\;
    \eqref{constraint_DCS}, \nonumber
\end{align}
where $a(t)$ denotes the joint action vector of all UAVs at time $t$, $E_i(t)$ is the energy consumption of drone $i$, and $\lambda>0$ is a trade-off parameter between QoS and energy. The weights and trade-off parameters are tuned empirically; the values used in the experiments are reported in Table~\ref{tab:hyperparameters}.

%%%%%%%%%%%%%%%%%%%%%%%%%%%%%%%%%%%%%%%%%%%%%%%%%%%%%%%%%%
\subsection{Metric computation: latency, throughput, and SINR}
\label{subsec:metric_computation}
%%%%%%%%%%%%%%%%%%%%%%%%%%%%%%%%%%%%%%%%%%%%%%%%%%%%%%%%%%

We briefly describe how latency, throughput, and SINR are computed in the simulator.

\paragraph*{SINR}
For each active drone–user link, the instantaneous SINR is
\begin{equation}
    \mathrm{SINR}_{i,m}(t)
    =
    \frac{P_i G_i G_m \, |h_{i,m}(t)|^2}{\displaystyle
    \sum_{k \neq i} P_k G_k G_m \, |h_{k,m}(t)|^2 + N_0},
\end{equation}
where $P_i$ is the transmit power of drone $i$, $G_i$ and $G_m$ are antenna gains, $h_{i,m}(t)$ is the complex channel coefficient (LOS/NLOS as in Section~\ref{Sec:mod}), and $N_0$ is the noise power. The values $S_i^X(t)$ in~\eqref{equation:qos_metric_updated} are obtained by averaging $\mathrm{SINR}_{i,m}(t)$ over users of type $X$ served by drone $i$.

\paragraph*{Throughput}
Throughput is obtained from SINR via a Shannon-like mapping:
\begin{equation}
    T_{i,m}(t)
    =
    B \, \log_2 \!\bigl( 1 + \mathrm{SINR}_{i,m}(t) \bigr),
\end{equation}
where $B$ is a fixed effective bandwidth per user (in MHz). In the implementation, we take $B=36$~MHz and use $T_{i,m}(t)$ as a proxy for achievable data rate in Mbps, implicitly assuming each active user is allocated an independent link of bandwidth $B$. The quantities $T_i^X(t)$ in~\eqref{equation:qos_metric_updated} and in Tables~\ref{Table:Rural_user_distribution}--\ref{Table:Urban_users_distribution} are averages of $T_{i,m}(t)$ over users of type $X$ served by drone $i$.

\paragraph*{Latency}
We do not implement an explicit queueing/scheduling model. Instead, latency is modeled as a proxy combining propagation and service components. For a drone–user link of distance $d_{i,m}(t)$ and throughput $T_{i,m}(t)$,
\begin{equation}
    L_{i,m}(t)
    =
    \Biggl(
        \frac{d_{i,m}(t)}{c}
        +
        \frac{1}{T_{i,m}(t) + \varepsilon}
    \Biggr)
    \bigl( 1 + \kappa \, v_i(t) \bigr),
\end{equation}
where $c$ is the speed of light, $v_i(t)$ is the speed of drone $i$, $\kappa$ is a small scaling factor, and $\varepsilon>0$ avoids division by zero. This expression is evaluated in seconds and then rescaled into the range $[10^{-3}, 4 \times 10^{-2}]$ for our QoS targets. The reported ``ms'' values in Tables~\ref{Table:Rural_user_distribution}--\ref{Table:Urban_users_distribution} thus act as normalized latency proxies rather than strict one-way delay measurements.

\paragraph*{Absolute vs.\ normalized units}
Both latency ($L_\text{target}^X$) and throughput ($T_\text{target}^X$) targets are design goals used inside the reward shaping, not hard service-level guarantees. The numerical values (e.g., $500$ ``Mbps'' for type~$A$) should be read as normalized per-user capacity benchmarks under the above mapping, not as exact physical link budgets under a fully specified MCS and scheduler.

%%%%%%%%%%%%%%%%%%%%%%%%%%%%%%%%%%%%%%%%%%%%%%%%%%%%%%%%%%
\section{MADRL algorithms}\label{Sec:MADRL}
%%%%%%%%%%%%%%%%%%%%%%%%%%%%%%%%%%%%%%%%%%%%%%%%%%%%%%%%%%
This section presents our approach to applying Multi-Agent Deep Reinforcement Learning (MADRL) in UAV-assisted wireless communication systems. These systems are characterized by dynamic user mobility, fluctuating traffic demands, and constrained energy resources, which require adaptive and efficient optimization strategies. We implement three MADRL algorithms—Multi-Agent Proximal Policy Optimization (MAPPO), Multi-Agent Deep Deterministic Policy Gradient (MADDPG), and Multi-Agent Deep Q-Network (MADQN)—selected for their ability to handle continuous or discrete action spaces in dynamic and uncertain environments.

The primary objective is to optimize UAV-based resource allocation, user coverage, and energy management, thereby ensuring robust QoS in terms of latency, throughput, and SINR. To achieve this, we simulate a realistic environment for UAV agents that captures the main aspects of mobility, traffic, and resource constraints.

%%%%%%%%%%%%%%%%%%%%%%%%%%%%%%%%%%%%%%%%%%%%%%%%%%%%%%%%%%
\subsection{Environment simulation algorithm}
%%%%%%%%%%%%%%%%%%%%%%%%%%%%%%%%%%%%%%%%%%%%%%%%%%%%%%%%%%
The simulation workflow, presented in Algorithm~\ref{Algo: Rural_environment}, begins by initializing the environment with grid layout parameters, user types ($A$, $B$, $C$), and QoS targets. UAVs are initially positioned randomly, while mobile users follow a random walk~\cite{g2021random}, simulating dynamic changes in location and traffic demand. To ensure realistic channel modeling, we adopt a 3GPP TR~38.901-inspired propagation model~\cite{etsi2018138}, incorporating path loss, outdoor-to-indoor (O2I) penetration, LoS/NLoS effects, shadowing, and Rician fading, as outlined in Section~\ref{Sec:mod}. These enhancements tailor the channel model for UAV-assisted 5G slicing and improve robustness under dynamic network conditions.

At each simulation step, UAVs execute actions to optimize coverage, maintain connectivity, and minimize energy consumption. Collision-avoidance mechanisms adjust UAV trajectories, while a battery management system directs low-energy UAVs to the dynamic charging station. QoS metrics such as latency, throughput, and SINR are evaluated in real time, enabling continuous UAV-action optimization through feedback from the MARL algorithm. Inter-UAV communication is enabled within the communication range $d_{\text{com}}$, allowing UAVs to coordinate, share user-location data, and prioritize high-demand type-$A$ users. The process continues for a predetermined number of steps, followed by an analysis of the resulting QoS metrics and energy consumption. The algorithm for the urban scenario is similar to Algorithm~\ref{Algo: Rural_environment}, with the specific parameters and constraints described in Section~\ref{Sec:mod}.

\begin{algorithm}
\caption{Environment simulation algorithm}\label{Algo: Rural_environment}
\begin{algorithmic}[1]
    \State \textbf{Initialize:} Environment parameters from Tables~\ref{Table:Rural_user_distribution}--\ref{Table:Urban_users_distribution}--\ref{tab:Simulationparametrs}, QoS targets, communication range $d_{\text{com}}$
    \While {simulation step $t \leq T$}
        \For {each drone $i = 1, \dots, N$}
            \State \textbf{Compute} QoS metrics (latency, throughput, SINR) for each user $m_i$ within range
            \State \textbf{Calculate} reward $r(t)$ using Algorithm~\ref{Algo:PPO}, Algorithm~\ref{Algo:DDPG}, or Algorithm~\ref{Algo:PriorityReplayDQN}
            \State \textbf{Movement planning:} update position based on user density and collision avoidance
            \If{$B_i(t) \leq B_{\min}$} 
                \State send drone $i$ to the nearest charging station
            \EndIf
        \EndFor
        \State \textbf{Update} charging-station coordinates: compute centroid of user positions
        \State \textbf{Update} user positions $u_m(t)$ using the (rural or urban) random-walk model
        \State \textbf{Communication:} drones within $d_{\text{com}}$ share information and prioritize type-$A$ users
    \EndWhile
    \State \textbf{Return} QoS metrics, reward performance, and coverage maps
\end{algorithmic}
\end{algorithm}

%%%%%%%%%%%%%%%%%%%%%%%%%%%%%%%%%%%%%%%%%%%%%%%%%%%%%%%%%%
\subsection{Overview of MARL}\label{Sec:Pre}
%%%%%%%%%%%%%%%%%%%%%%%%%%%%%%%%%%%%%%%%%%%%%%%%%%%%%%%%%%
Reinforcement learning (RL) involves learning optimal actions through interactions with an environment to maximize cumulative rewards. RL problems are often modeled as Markov decision processes (MDPs), where the future depends only on the current state and action. In multi-agent reinforcement learning (MARL), multiple autonomous agents interact in a shared environment and learn policies while accounting for the actions of others~\cite{albrecht2024multi}. This setting is particularly relevant to robotics and telecommunications, where cooperative or competitive interactions are common. Typical MARL paradigms include:
\begin{itemize}
    \item Independent learning (IL): each agent learns its policy independently, which can lead to non-stationarity from the perspective of any single agent.
    \item Centralized training with decentralized execution (CTDE): agents use global information during training but act independently based on local observations during execution.
    \item Cooperative MARL: agents work together to optimize shared goals.
    \item Mixed cooperative–competitive MARL: agents may either cooperate or compete (e.g., in traffic management or multi-agent trading).
\end{itemize}
In this work, we adopt the CTDE approach, which provides a good compromise between sample efficiency during training and practicality during real-time execution.

%%%%%%%%%%%%%%%%%%%%%%%%%%%%%%%%%%%%%%%%%%%%%%%%%%%%%%%%%%
\subsection{Centralized training with decentralized execution}
\begin{figure}[htbp]
    \centering
    \includegraphics[width=0.4\textwidth]{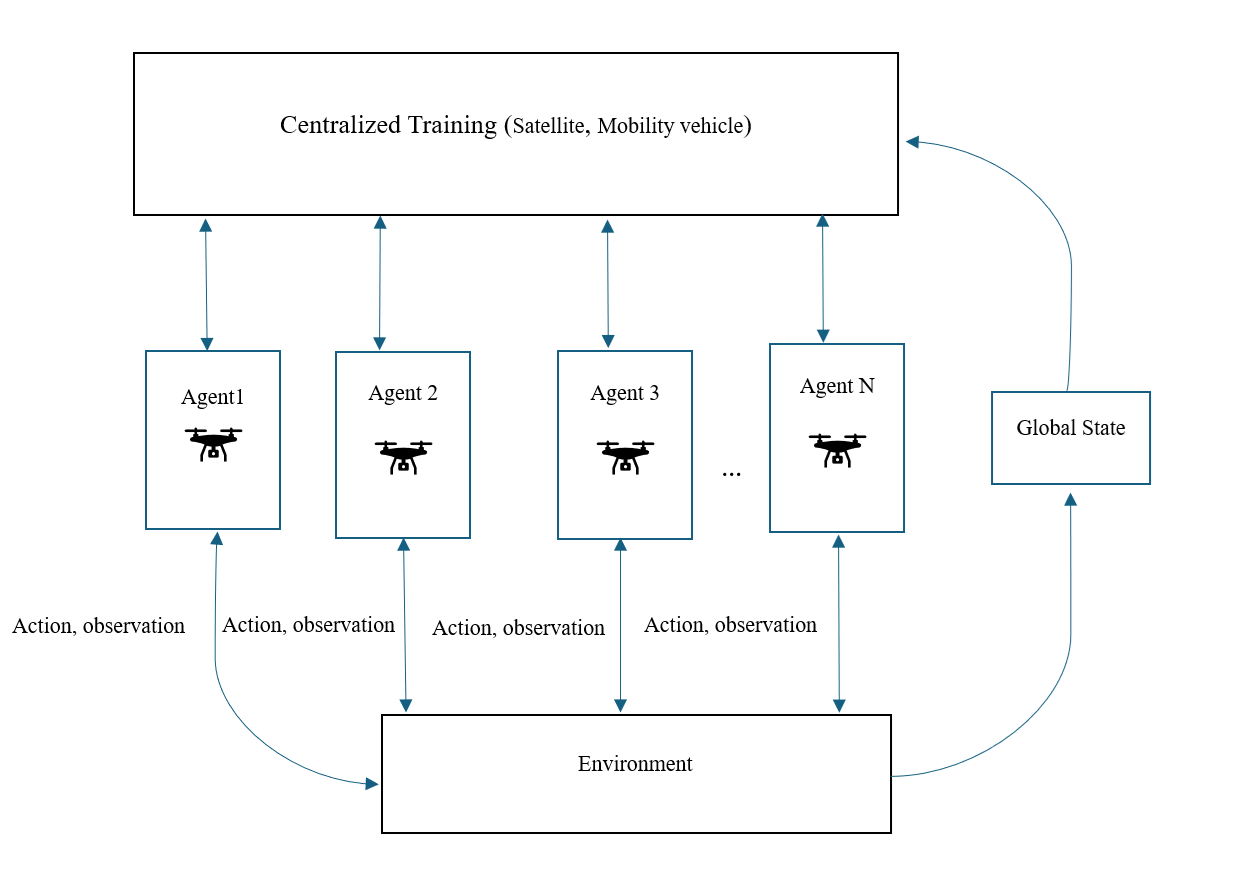}
    \caption{CTDE (centralized training, decentralized execution) concept in a multi-agent setting.}
    \label{fig:CTDE}
\end{figure}

%%%%%%%%%%%%%%%%%%%%%%%%%%%%%%%%%%%%%%%%%%%%%%%%%%%%%%%%%%
In our framework, the CTDE paradigm enables coordinated strategy learning during training while allowing independent decision making during deployment. During centralized training, UAVs can access global state information and learn strategies that optimize network-wide objectives such as throughput and SINR. During execution, each UAV operates autonomously using only local observations, as illustrated in Figure~\ref{fig:CTDE}.

\subsubsection{Centralized training phase}
During centralized training, all UAV agents have access to the global state $X(t)$, defined by aggregating the local observations $x_i(t)$ from each UAV:
\begin{equation}
X(t) = [x_1(t), x_2(t), \dots, x_N(t)],
\end{equation}
where $x_i(t)$ denotes the local observation of UAV $i$ at time $t$, and $N$ is the number of UAVs. The centralized critic learns the joint action-value function $Q(X, a_1, a_2, \dots, a_N;\phi)$ by minimizing the loss:
\begin{align}
L(\phi) = \mathbb{E} \Big[ \big(& r + \gamma\, Q\big(X(t+1), a'_1,\dots,a'_N;\phi'\big) \nonumber \\[-2mm]
& - Q\big(X(t), a_1,\dots,a_N;\phi\big) \big)^2 \Big],
\end{align}
where $\phi$ and $\phi'$ are the parameters of the critic and its target network, respectively, $r$ is the immediate reward, and $\gamma$ is the discount factor. Centralized training allows the critic to capture interdependence's among UAVs and thus improves the stability and convergence of the learned policies.

\subsubsection{Decentralized execution under partial observability}
During real-time deployment, global state information is unavailable, and each UAV must independently select actions based on its local observation $x_i(t)$. Each agent employs a decentralized policy $\pi_i$ parameterized by $\theta_i$, mapping local states to actions:
\begin{equation}
a_i(t) = \pi_i(x_i(t);\theta_i).
\end{equation}
Policy parameters are updated using
\begin{equation}
\nabla_{\theta_i} J(\theta_i) = \mathbb{E}\left[\nabla_{\theta_i}\log\pi_i(a_i|x_i;\theta_i)\,A_i(X,a_i)\right],
\end{equation}
where the advantage function $A_i(X,a_i)$ is computed by
\begin{equation}
A_i(X,a_i) = Q(X,a_1,\dots,a_N) - V(X),
\end{equation}
with $V(X)$ serving as a baseline value for variance reduction. This structure ensures that each UAV’s decentralized policy implicitly accounts for the global conditions encountered during training, which mitigates partial observability and enables robust real-time decision making. We consider two environments, rural and urban, both designed to simulate UAV–user interactions under the QoS metrics of latency, throughput, and SINR.

\paragraph*{CTDE across algorithms.}
MAPPO and MADDPG implement CTDE in the standard way: a centralized critic
receives the joint observation and joint actions during training, while each
actor relies only on its local observation during execution. For MADQN, the
environment provides all UAVs with the same global observation vector; thus,
MADQN uses parameter sharing and a shared reward signal without a centralized
critic. Since the observation available at execution is identical to training,
the CTDE assumption is satisfied in a trivial sense for MADQN.

\subsection{State representation and partial observability}
\label{subsec:state_representation}

Each UAV agent receives a fixed-length observation vector constructed by the simulator at every time step. For $N$ drones, the observation is
\begin{equation}
    x(t) \in \mathbb{R}^{d_s},
\end{equation}
with $d_s = 6N + 12$. In all experiments we set $N=4$, which yields $d_s = 36$. The components of $x(t)$ are:

\begin{itemize}
    \item \textbf{Drone positions}:
    the $(x,y,z)$ coordinates of all drones, clipped to the admissible region
    $[0,3] \times [0,3] \times [h_{\min}, h_{\max}]$ for the urban scenario and
    $[0,10] \times [0,10] \times [h_{\min}, h_{\max}]$ for the rural scenario.
    This contributes $3N$ scalars.

    \item \textbf{Battery levels}:
    the normalized remaining battery life of each drone $i$, denoted $B_i(t) \in [0,100]$ (percentage).
    This contributes $N$ scalars.

    \item \textbf{User counts}:
    the current numbers of associated users of each priority type, $(n_A(t),n_B(t),n_C(t))$, aggregated over all drones.
    This contributes $3$ scalars.

    \item \textbf{Drone velocities}:
    the instantaneous 3D velocity vectors of all drones, $\mathbf{v}_i(t)$, contributing another $3N$ scalars.

    \item \textbf{Aggregated QoS metrics}:
    for each user type $X\in\{A,B,C\}$, we include the current averaged latency $L^X(t)$, throughput $T^X(t)$, and SINR $S^X(t)$ across all users of type $X$ in the system.
    This contributes $3 \times 3 = 9$ scalars.
\end{itemize}
Thus, for $N=4$ drones, the total observation dimension is
\[
d_s = 3N + N + 3 + 3N + 9 = 36.
\]

%%%%%%%%%%%%%%%%%%%%%%%%%%%%%%%%%%%%%%%%%%%%%%%%%%%%%%%%%%
\subsection{MADRL algorithms}
%%%%%%%%%%%%%%%%%%%%%%%%%%%%%%%%%%%%%%%%%%%%%%%%%%%%%%%%%%
By incorporating deep neural networks, MARL becomes MADRL, which enhances agents’ ability to handle large state spaces and learn complex policies in dynamic environments~\cite{albrecht2024multi}. We leverage MADRL to develop scalable solutions in our multi-UAV system, focusing on MAPPO, MADDPG, and MADQN to optimize QoS in the drone network. The corresponding hyperparameters are summarized in Table~\ref{tab:hyperparameters}.

%%%%%%%%%%%%%%%%%%%%%%%%%%%%%%%%%%%%%%%%%%%%%%%%%%%%%%%%%%
\subsection{Simulation environment and scenario details}
%%%%%%%%%%%%%%%%%%%%%%%%%%%%%%%%%%%%%%%%%%%%%%%%%%%%%%%%%%
To evaluate the performance of each MADRL algorithm, we consider both rural and urban environments.

\subsubsection{Rural environment}
A $10\times 10$ grid represents the rural environment, modeling a sparsely populated area with the following characteristics:
\begin{itemize}
  \item Grid size: $10 \times 10$ units.
  \item User distribution: users are categorized into three priority types $(A, B, C)$, each with distinct QoS requirements; type~$A$ users have the highest priority.
  \item UAV constraints: UAVs are initialized randomly with a communication range of $5$ units and manage battery life by returning to a charging station when energy falls below a threshold.
  \item QoS metrics: latency, throughput, and SINR are used to capture the challenges of user sparsity and mobility.
\end{itemize}
The numerical parameters are summarized in Table~\ref{Table:Rural_user_distribution}.

\subsubsection{Urban environment}
The urban environment is modeled as a dense $3\times 3$ grid with higher user density and stronger signal obstructions:
\begin{itemize}
\item Grid size: $3 \times 3$ units, representing a compact dense area.
\item User distribution: higher user density with the same three-tier prioritization $(A, B, C)$ as in the rural case. UAVs operate with a communication range of one unit due to the interference-rich environment.
\item UAV constraints: UAVs perform more frequent adjustments to maintain connectivity under strong interference and NLoS effects.
\item QoS metrics: emphasis is placed on reducing latency and energy consumption while respecting user-priority requirements.
\end{itemize}
The corresponding parameters are summarized in Table~\ref{Table:Urban_users_distribution}.

\begin{table}[htbp]
\centering
\caption{QoS targets and user parameters in the rural environment}
\label{Table:Rural_user_distribution}
\scriptsize
\begin{tabular}{|l|c|c|c|}
\hline
\rowcolor{lightgray} \textbf{Metric} & \textbf{Type A} & \textbf{Type B} & \textbf{Type C} \\
\hline
Number of users & 1--90 & 1--50 & 1--40 \\
\hline
Target latency (ms) & 1  & 10 & 40 \\
\hline
Target throughput (Mbps) & 500 & 350 & 200 \\
\hline
Target SINR (dB) & 25 & 15 & 10 \\
\hline
Communication range (units) & \multicolumn{3}{c|}{5.0} \\
\hline
Coverage range (units)      & \multicolumn{3}{c|}{3.0} \\
\hline
Grid size (units)           & \multicolumn{3}{c|}{$10 \times 10$} \\
\hline
\end{tabular}
\end{table}

\begin{table}[htbp]
\centering
\caption{QoS targets and user parameters in the urban environment}
\label{Table:Urban_users_distribution}
\scriptsize
\begin{tabular}{|l|c|c|c|}
\hline
\rowcolor{lightgray} \textbf{Metric} & \textbf{Type A} & \textbf{Type B} & \textbf{Type C} \\
\hline
Number of users & 1--90 & 1--50 & 1--40 \\
\hline
Target latency (ms) & 1  & 10 & 40 \\
\hline
Target throughput (Mbps) & 500 & 350 & 200 \\
\hline
Target SINR (dB) & 25 & 15 & 10 \\
\hline
Communication range (units) & \multicolumn{3}{c|}{2.0} \\
\hline
Coverage range (units)      & \multicolumn{3}{c|}{1.5} \\
\hline
Grid size (units)           & \multicolumn{3}{c|}{$3 \times 3$} \\
\hline
\end{tabular}
\end{table}

%%%%%%%%%%%%%%%%%%%%%%%%%%%%%%%%%%%%%%%%%%%%%%%%%%%%%%%%%%
\subsection*{Modeling choices, abstractions, and limitations}
%%%%%%%%%%%%%%%%%%%%%%%%%%%%%%%%%%%%%%%%%%%%%%%%%%%%%%%%%%

We summarize here several modeling choices that were intentionally simplified to
keep the simulator lightweight and comparable across MAPPO, MADDPG, and
MADQN.

\paragraph*{QoS targets and metric units.}
The throughput and latency targets in Tables~\ref{Table:Rural_user_distribution}--\ref{Table:Urban_users_distribution}
(e.g., 500/350/200~Mbps and 1/10/40~ms) are \emph{scenario-level normalized goals} rather than strict NR-layer guarantees. 
Throughput is mapped from SINR via a Shannon-like formula with a fixed effective bandwidth of 36~MHz, and latency is a proxy combining propagation and service delay (see Section~IV-D). 
The reported ``ms'' and ``Mbps'' values should therefore be interpreted as normalized service levels inside the reward, used to compare policies consistently across scenarios. 
The target values are inspired by typical 5G/NR service categories: type-A users correspond to high-demand eMBB slices, type-B users to mid-tier interactive services, and type-C users to background/best-effort traffic. 
This abstraction follows common practice in UAV–5G MARL studies~\cite{messaoud2020deep,kim2024cooperative,tarekegn2022deep}, where the focus is on the \emph{relative} behavior of learning algorithms under heterogeneous demands.

\paragraph*{Spatial units.}
All coordinates use normalized spatial units instead of meters. A $3\times 3$
``urban'' grid may correspond to a few hundred meters, whereas a $10\times 10$
rural grid can represent a larger macro-cell area. Only relative distances enter
the TR~38.901-inspired channel model; conversion to meters is left for
deployment studies. This ensures consistency between the grids in
Tables~\ref{Table:Rural_user_distribution}--\ref{Table:Urban_users_distribution}
and the channel formulation in Section~III.

\paragraph*{Constraint handling.}
Operational constraints (altitude limits, collision avoidance, communication
range, disjoint service sets) are enforced through reward penalties and simple
heuristics. The framework does not implement a CMDP or control-barrier-function
mechanism; constraints are encouraged in expectation rather than guaranteed at
execution time. This design is appropriate for a comparative MARL study and avoids
additional algorithmic complexity that is orthogonal to the main objective.

\paragraph*{Energy model.}
The energy model uses a normalized battery percentage with hover cost,
velocity-proportional cost, and per-user service cost. The mobile charging station
represents an abstract service facility (e.g., a relocatable ground charger) without an explicit relocation-cost model. 
These simplifications highlight QoS–energy trade-offs while avoiding platform-specific mechanical modeling, which would
complicate MARL comparisons without changing the qualitative conclusions.

\paragraph*{Discount factors.}
The discount choices (MAPPO: $0.99$, MADDPG: $1.0$, MADQN: $0.01$) were
empirically tuned for stability. MADQN uses a very small $\gamma$ to avoid
unstable Q-value growth in a highly non-stationary multi-agent setting.
MADDPG benefits from an undiscounted episodic return in our finite-horizon tasks, while MAPPO uses a
standard long-horizon value. These choices were tested for robustness and do not
change the qualitative performance ordering of the algorithms.

%%%%%%%%%%%%%%%%%%%%%%%%%%%%%%%%%%%%%%%%%%%%%%%%%%%%%%%%%%
\subsection{Algorithm design in dynamic environments}
%%%%%%%%%%%%%%%%%%%%%%%%%%%%%%%%%%%%%%%%%%%%%%%%%%%%%%%%%%
We now describe how the three MADRL algorithms—MAPPO, MADDPG, and MADQN—are instantiated in our UAV environment. The hyperparameters for each algorithm are listed in Table~\ref{tab:hyperparameters}.

\subsubsection{MAPPO}
Multi-agent PPO (MAPPO) is used for continuous fine-tuning of UAV positions and other control variables in dynamic environments to optimize QoS. Algorithm~\ref{Algo:PPO} outlines the MAPPO procedure, where each UAV employs an actor–critic model to select actions based on observed states, including user demand, channel conditions, and energy levels. Policy updates rely on the PPO clipped objective, which helps stabilize training in highly dynamic scenarios.

\begin{algorithm}[htbp]
\caption{MAPPO for UAV-based QoS optimization}
\label{Algo:PPO}
\begin{algorithmic}[1]
\State \textbf{Initialize} environment of Algorithm~\ref{Algo: Rural_environment}, rollout buffer, and MAPPO hyperparameters from Table~\ref{tab:hyperparameters}
\State Reset environment and observe initial state $s_0$
\For{each time step $t = 1$ to $\texttt{T}$}
        \State Select joint action $a(t)$ using policy $\pi_{\theta}(s(t))$
        \State Execute $a(t)$ in the environment, observe $s(t+1)$, reward $r(t)$, and done flag $d$
        \State Store transition $(s(t), a(t), r(t), s(t+1), d)$ in the rollout buffer
        \If{the buffer is full}
            \State Sample mini-batches, compute advantages, and update actor–critic models
        \EndIf
        \If{$d$ is true}
            \State end episode and store the total return $R_e$
        \EndIf
\EndFor
\State \textbf{Return} trained actor–critic models
\end{algorithmic}
\end{algorithm}

\subsubsection{MADDPG}
Algorithm~\ref{Algo:DDPG} implements Multi-Agent Deep Deterministic Policy Gradient (MADDPG) for continuous control of UAV positions and power levels. Each UAV uses an $\epsilon$-greedy exploration strategy around its deterministic policy. Transitions are stored in a replay buffer and used to update the actor–critic networks, with soft updates applied to the target networks to maintain stability.

\begin{algorithm}[htbp]
\caption{MADDPG for UAV-based QoS optimization}
\label{Algo:DDPG}
\begin{algorithmic}[1]
\State \textbf{Initialize} environment of Algorithm~\ref{Algo: Rural_environment}, replay buffer, and MADDPG hyperparameters from Table~\ref{tab:hyperparameters}
\State Reset environment and observe initial state $s_0$
\For{each step $t=1$ to $\texttt{T}$}
        \State Select joint action $a(t)$ using an $\epsilon$-greedy policy around the current actors
        \State Execute $a(t)$, observe next state $s(t+1)$, reward $r(t)$, and done flag $d$
        \State Store transition $(s(t), a(t), r(t), s(t+1), d)$ in the replay buffer
        \If{the buffer is full}
            \State Sample a mini-batch, compute TD targets, and update actor–critic networks
            \State Soft-update the target networks
        \EndIf
        \If{$d$ is true}
            \State break the episode
        \EndIf
        \State Log rewards and QoS metrics, adjust the exploration rate $\epsilon$
\EndFor
\State \textbf{Return} trained actor–critic models
\end{algorithmic}
\end{algorithm}

\subsubsection{MADQN}
Algorithm~\ref{Algo:PriorityReplayDQN} applies Multi-Agent Deep Q-Networks (MADQN) to a discretized action space, suitable for scenarios with predefined UAV positions or power levels. The action space is discretized to approximate continuous actions while retaining computational efficiency. A similar discretization strategy has been used in~\cite{messaoud2020deep,kim2024cooperative,tarekegn2022deep}. The multi-dimensional action set $\mathcal{A}$ is constructed as
\begin{equation}
    \mathcal{A} = \texttt{Meshgrid}\left( \left\{ a_{\text{min}} + k \Delta a : k = 0, \ldots, N_p-1 \right\} \right)^{d_a},
\end{equation}
where $\Delta a$ is the discretization step size, $a_{\min}$ is the minimum value in the action space, $N_p$ is the number of discrete points per dimension, and $d_a$ is the action-space dimension. The operator \texttt{Meshgrid} denotes the standard procedure for generating a grid of points from discrete values in each dimension.

\begin{algorithm}[htbp]
\caption{MADQN with discretized action space for UAV-based QoS optimization}
\label{Algo:PriorityReplayDQN}
\begin{algorithmic}[1]
\State \textbf{Initialize} environment, replay buffer $\mathcal{D}$, hyperparameters, Q-network, and target network
\State \textbf{Discretize} action space $\mathcal{A}$, initialize exploration factor $\epsilon = 1.0$
\For{each episode}
    \State Reset environment and observe initial state $s_0$
    \For{each time step $t$}
        \State Select action $a_t \in \mathcal{A}$ using an $\epsilon$-greedy policy
        \State Execute $a_t$, observe reward $r_t$, next state $s_{t+1}$, and done flag $d_t$
        \State Store transition $(s_t, a_t, r_t, s_{t+1}, d_t)$ in $\mathcal{D}$
        \If{the buffer is full}
            \State Sample a mini-batch, compute target Q-values, update the Q-network
            \State Soft-update the target network and update $\epsilon$
        \EndIf
        \If{$d_t$ is true}
            \State end episode and log results
        \EndIf
    \EndFor
\EndFor
\State \textbf{Return} trained Q-network and optimized QoS metrics
\end{algorithmic}
\end{algorithm}

%%%%%%%%%%%%%%%%%%%%%%%%%%%%%%%%%%%%%%%%%%%%%%%%%%%%%%%%%%
\subsubsection{Reward function for QoS optimization}
%%%%%%%%%%%%%%%%%%%%%%%%%%%%%%%%%%%%%%%%%%%%%%%%%%%%%%%%%%
A key component of our approach is the design of the reward function, which aligns directly with QoS metrics (latency, throughput, SINR) and energy consumption. The reward structure incentivizes drones to balance multiple objectives while optimizing performance and minimizing operational costs, as in~\cite{bista2024differentiated}.

\paragraph{Latency penalty}
UAVs incur penalties for excessive latency, particularly for high-priority type-$A$ users:
\begin{align}
    r_{\text{latency}}(t) = -\sum_{i=1}^{N} \Big( 
    & \omega_A \frac{L_i^A(t) - L_{\text{target}}^A}{L_{\text{target}}^A} \nonumber \\ 
    & + \omega_B \frac{L_i^B(t) - L_{\text{target}}^B}{L_{\text{target}}^B} \nonumber \\  
    & + \omega_C \frac{L_i^C(t) - L_{\text{target}}^C}{L_{\text{target}}^C} \Big),
    \label{equation:latency}
\end{align}
where $L_i^X(t)$ and $L_{\text{target}}^X$ are the experienced and target latencies for user type $X \in \{A,B,C\}$ served by drone $i$. The weights $\omega_A > \omega_B > \omega_C$ match those in~\eqref{equation:optimisation}. Negative contributions are applied when $L_i^X(t) > L_{\text{target}}^X$.

\paragraph{Throughput reward}
Positive rewards are granted when throughput meets or exceeds target thresholds:
\begin{align}
       r_{\text{throughput}}(t) = \sum_{i=1}^{N} \Big( 
       & \omega_A \frac{T_i^A(t) - T_{\text{target}}^A}{T_{\text{target}}^A} \nonumber \\ 
       & + \omega_B \frac{T_i^B(t) - T_{\text{target}}^B}{T_{\text{target}}^B} \nonumber \\  
       & + \omega_C \frac{T_i^C(t) - T_{\text{target}}^C}{T_{\text{target}}^C} \Big),
       \label{equation:Throughput}
\end{align}
where $T_i^X(t)$ and $T_{\text{target}}^X$ denote the experienced and target throughput for user type $X$.

\paragraph{SINR reward}
Similarly, UAVs are rewarded for maintaining acceptable SINR levels:
\begin{align}
    r_{\text{SINR}}(t) = \sum_{i=1}^{N} \Big( 
    & \omega_A \frac{S_i^A(t) - S_{\text{target}}^A}{S_{\text{target}}^A} \nonumber \\ 
    & + \omega_B \frac{S_i^B(t) - S_{\text{target}}^B}{S_{\text{target}}^B} \nonumber \\  
    & + \omega_C \frac{S_i^C(t) - S_{\text{target}}^C}{S_{\text{target}}^C} \Big),
    \label{equation:SINR}
\end{align}
where $S_i^X(t)$ and $S_{\text{target}}^X$ represent the experienced and target SINR for user type $X$.

\paragraph{Energy-related terms}
To encourage efficient UAV operation, we incorporate several energy-related terms. The efficiency reward is defined as
\begin{equation}
    r_{\text{eff}}(t) = - \frac{\sum_{i=1}^{N} \|v_i(t)\|}{\sum_{i=1}^{N} \left(E_{\text{max}} - E_i(t)\right) + 10^{-5}},
\end{equation}
where $N$ is the number of drones, $\|v_i(t)\|$ is the velocity norm of drone $i$, $E_{\text{max}}$ is the initial battery capacity, and $E_i(t)$ is the remaining battery level of drone $i$ at time $t$. A small constant is added to avoid division by zero.

To prioritize type-$A$ users, we add a coverage bonus:
\begin{equation}
    r_{\text{bonus}}(t) = 5 \, n_{A}(t),
\end{equation}
where $n_{A}(t)$ is the number of type-$A$ users within the UAVs’ coverage at time $t$.

To prevent UAV failures, a low-battery penalty $P_{\text{low}}(t)$ is applied:
\begin{equation}
 P_i^{\text{low}}(t) =
    \begin{cases} 
        -5 & \text{if } E_i(t) < 10 \text{ and serving type-$A$ users}, \\[0.5mm]
        -10 & \text{if } E_i(t) < 10 \text{ and not serving type-$A$ users}, \\[0.5mm]
        0 & \text{otherwise},
    \end{cases}
    \label{equation:Plow}
\end{equation}
and
\begin{equation}
P_{\text{low}}(t) = \sum_{i=1}^{N}  P_i^{\text{low}}(t).
\end{equation}

The total energy expenditure is modeled as
\begin{equation}
    E_{\text{total}}(t) = \sum_{i=1}^{N} \left( 0.5 \, \|v_i(t)\| + E_i^{\text{hover}}(t) + n_i^{\text{users}} \, E_i^{\text{user}}(t) \right),
\end{equation}
where $E_i^{\text{hover}}(t)$ is the hovering energy rate of drone $i$, $E_i^{\text{user}}(t)$ is the per-user energy rate, and $n_i^{\text{users}}$ is the number of users served by drone $i$.

The combined energy-related reward is
\begin{equation}
    r_{\text{energy}}(t) = r_{\text{eff}}(t) + r_{\text{bonus}}(t) - E_{\text{total}}(t) + P_{\text{low}}(t).
\end{equation}

Finally, the total reward at time $t$ is given by
\begin{eqnarray}
    r(t) &=& R_w \left( r_{\text{latency}}(t) + r_{\text{throughput}}(t) 
    + r_{\text{SINR}}(t) \right) \nonumber\\
    &&+ E_w \, r_{\text{energy}}(t),
    \label{equation:Totalrewards}
\end{eqnarray}
where $R_w$ and $E_w$ are scalar weights controlling the trade-off between QoS and energy. By tuning these weights (see Table~\ref{tab:hyperparameters}), the system can prioritize QoS metrics in accordance with user needs while maintaining energy efficiency. This reward formulation is applied at the agent level and drives UAV behavior toward the overall system objectives.

%%%%%%%%%%%%%%%%%%%%%%%%%%%%%%%%%%%%%%%%%%%%%%%%%%%%%%%%%%
\section{Simulation Setup and Evaluation}\label{Sec:simu}
%%%%%%%%%%%%%%%%%%%%%%%%%%%%%%%%%%%%%%%%%%%%%%%%%%%%%%%%%%
This section presents the simulation setup and results used to evaluate the performance of the proposed MADRL algorithms (MAPPO, MADDPG, and MADQN) in a UAV-based wireless communication environment. The objective is to optimize QoS for user types $A$, $B$, and $C$, while minimizing energy consumption and ensuring continuous coverage. The simulations were conducted using Python~3.7.8 and TensorFlow~2.11.0 on a system with an 11th Gen Intel Core i7-11850H CPU at 2.50GHz and 32~GB RAM (64-bit OS).

We simulate a 3D environment where multiple drones serve as mobile base stations, providing wireless coverage to ground users grouped by priority as described in Sections~\ref{Sec:mod} and~\ref{Sec:optim}. The simulation environment (channel model, user mobility, energy model) remains identical across all experiments and is applied uniformly to the three MADRL algorithms in both rural and urban scenarios.

\begin{table}[ht]
\centering
\caption{Units and conventions used in all metrics}
\begin{tabular}{l l}
\hline
\textbf{Metric} & \textbf{Unit / convention} \\
\hline
Latency & milliseconds (ms, normalized proxy) \\
Throughput & Megabits per second (Mbps, normalized) \\
SINR & decibels (dB) \\
Energy consumption & Battery percentage per step (\%) \\
Grid coordinates & Normalized spatial units (dimensionless) \\
Drone power & Normalized (0--1) \\
\hline
\end{tabular}
\label{tab:units}
\end{table}

%%%%%%%%%%%%%%%%%%%%%%%%%%%%%%%%%%%%%%%%%%%%%%%%%%%%%%%%%%
\subsection{Simulation parameters}
%%%%%%%%%%%%%%%%%%%%%%%%%%%%%%%%%%%%%%%%%%%%%%%%%%%%%%%%%%
The main simulation parameters are summarized in Table~\ref{tab:Simulationparametrs}. The environment features multiple UAVs, each controlled by a MADRL algorithm, optimizing their deployment to serve users efficiently while minimizing energy use.

\begin{table}[htbp]
	\centering
	\caption{Simulation parameters}
	\label{tab:Simulationparametrs}
	\small
	\begin{tabular}{ll}
		\hline\hline
		\textbf{Parameter} & \textbf{Value} \\
		\hline
		Number of drones & 4 \\
		User categories & type A, type B, type C \\
		User speed & Randomized in [1, 3] m/s \\
		Drone height limits & $h_d^{\min} = 1$, $h_d^{\max} = 10$ units \\
		Simulation time & 1000 time steps \\
		Initial charging station & [2.5, 1.0, 1.0] \\
		Hovering consumption & 0.1 per time step \\
		Full charge level & 90\% \\
		Charging threshold & 10\% \\
		Battery recovery rate & 2\% per step \\
		Maximum steps & 1000 \\
		Action space & Continuous / discrete \\
		Time step $\Delta t$ & 1.0 \\
		Max velocity $v_{\max}$ & 1.0 unit/step \\
		Battery capacity $E_{\max}$ & 100 units \\
		Battery threshold $B_{\min}$ & 10 units \\
		Station update limit & 1.0 unit/step \\
		Discount factor $\gamma$ & 0.99 \\
		Reward weights & See Table 2 \\
		\hline\hline
	\end{tabular}
\end{table}

Each algorithm's hyperparameters are listed in Table~\ref{tab:hyperparameters}. MAPPO and MADDPG operate in continuous action spaces, whereas MADQN uses a discretized action set. The learning rate is fixed at $0.001$, and discount factors are tuned per algorithm as detailed in Section~\ref{subsec:discount_factors}. Key parameter values (e.g., the 2\% battery-recovery rate per step) were empirically derived through iterative refinement, adjusting energy weights in the reward function and differentiating hover and flight energy costs. Although we do not present a formal Pareto front, the tuning methodology aligns with best practices from prior works~\cite{tarekegn2022deep, luong2021deep, abualigah2021applications, dilshad2020applications} that trade off UAV energy consumption against communication efficiency.

For MADQN and MADDPG, we use an experience replay buffer of size 50{,}000 with prioritized replay (exponent $\alpha=0.6$) and importance sampling. MAPPO, in contrast, is implemented in a strictly on-policy fashion: it relies on short-horizon rollout buffers that are discarded after each update and does \emph{not} use prioritized replay.

Exploration decay rates and target-network update intervals are configured to ensure training stability and adaptability in dynamic conditions. UAV control hyperparameters (Table~\ref{tab:hyperparameters}) are derived from prior studies to support robust adaptive positioning and multi-slice resource allocation.

\paragraph{Interpretation of spatial units}
In both urban and rural environments, the simulator operates using normalized
spatial units rather than physical meters. This avoids restricting MARL
training to a particular geographic scale. A $3\times 3$ unit ``urban'' grid
may correspond, for example, to a 300$\times$300~m area, while the $10\times 10$
unit rural grid may represent a larger macro-cell region such as 1~km$^2$.
Only relative distances matter for the RL dynamics and the TR~38.901-inspired
channel model; the mapping to meters can be chosen freely for deployment
studies.

\paragraph*{On-policy nature of MAPPO and replay buffer usage}
We acknowledge that the initial version of Table~\ref{tab:hyperparameters}
could be misinterpreted regarding MAPPO's data management. In our
implementation, MAPPO is strictly on-policy: at each update, we collect fresh
trajectories under the current policy, store them in a short-horizon rollout
buffer, and perform a fixed number of gradient steps. This buffer does not
accumulate off-policy experience across many policy updates and therefore
does not constitute a replay buffer in the usual sense.

Prioritized experience replay is used only for MADQN (and, where applicable,
MADDPG), where it is compatible with the off-policy learning paradigm.

\begin{table*}[htbp]
\centering
\caption{Hyperparameters for MAPPO, MADDPG, and MADQN algorithms}
\begin{tabular}{|l|c|c|c|}
\hline
\rowcolor{lightgray} \textbf{Parameter}      & \textbf{MAPPO}     & \textbf{MADDPG}    & \textbf{MADQN} \\ \hline
$\gamma$: discount factor      & 0.99            & 1.0           & 0.01 \\ \hline
$\tau$: soft update rate       & 0.005           & 0.01          & 0.01 \\ \hline
$\epsilon$: exploration rate   & 0.2             & 0.1           & 1.0 (min 0.5, decay 0.9) \\ \hline
Loss tracking                  & Actor and critic & Actor and critic & Critic \\ \hline
Actor-network size [neurons, neurons]  & \multicolumn{3}{c|}{[128, 128]}    \\ \hline
Critic-network size [neurons, neurons] & \multicolumn{3}{c|}{[128, 128]}   \\ \hline
Reward history tracking        & \multicolumn{3}{c|}{Yes} \\ \hline
Battery level monitoring       & \multicolumn{3}{c|}{Yes} \\ \hline
Reward weight $\omega_A$       & \multicolumn{3}{c|}{4.5} \\ \hline
Reward weight $\omega_B$       & \multicolumn{3}{c|}{2.5} \\ \hline
Reward weight $\omega_C$       & \multicolumn{3}{c|}{1.5} \\ \hline
Energy weight $E_w$            & \multicolumn{3}{c|}{1.0} \\ \hline
QoS weight $R_w$               & \multicolumn{3}{c|}{4.0} \\ \hline
Buffer size (MADDPG/MADQN)     & \multicolumn{3}{c|}{50{,}000} \\ \hline
Batch size                     & \multicolumn{3}{c|}{128} \\ \hline
State dimension                & \multicolumn{3}{c|}{36} \\ \hline
Action dimension               & \multicolumn{3}{c|}{3} \\ \hline
$\eta$: learning rate          & \multicolumn{3}{c|}{0.001} \\ \hline
$\alpha$: priority-replay exponent (MADDPG/MADQN) & \multicolumn{3}{c|}{0.6} \\ \hline
$\texttt{E}$: total number of episodes    & \multicolumn{3}{c|}{1000} \\ \hline
\end{tabular}
\label{tab:hyperparameters}
\end{table*}

\subsection{Discount factors and horizon interpretation}
\label{subsec:discount_factors}

The discount factors used in our experiments are summarized in Table~\ref{tab:hyperparameters}: $\gamma=0.99$ for MAPPO, $\gamma=1.0$ for MADDPG, and $\gamma=0.01$ for MADQN. We acknowledge that these choices are unconventional and clarify their motivation here.

For MAPPO, we adopt $\gamma=0.99$, which corresponds to a standard long-horizon setting over hundreds of simulation steps and is consistent with prior work on continuous-control MARL.

For MADDPG, we use $\gamma=1.0$ to emphasize cumulative performance over the full episode without discounting. In our episodic tasks with finite horizon, this choice did not lead to divergence in practice and produced stable learning in preliminary experiments. It should, however, be interpreted as a modeling decision favoring long-term energy management, rather than a recommendation for all continuous-control problems.

For MADQN, we set $\gamma=0.01$ to focus almost entirely on immediate rewards. This effectively turns MADQN into a myopic controller optimizing one-step QoS and energy proxies in a highly non-stationary multi-agent environment. During early tuning, larger values of $\gamma$ (e.g., $0.9$--$0.99$) led to unstable Q-value estimates and degraded performance in our setting. We therefore retain $\gamma=0.01$ as a pragmatic stabilization choice for this particular environment.

A full ablation over $\gamma$ values for all algorithms is left as future work. Consequently, the absolute performance rankings reported in Section~\ref{Sec:results} should be interpreted with the understanding that they reflect one specific, empirically tuned discount configuration rather than a globally optimized choice.

%%%%%%%%%%%%%%%%%%%%%%%%%%%%%%%%%%%%%%%%%%%%%%%%%%%%%%%%%%

%%%%%%%%%%%%%%%%%%%%%%%%%%%%%%%%%%%%%%%%%%%%%%%%%%%%%%%%%%
\subsection{Results and Comparative Analysis}
\label{Sec:results}
% Comparative analysis of MADRL algorithms in urban and rural environments

This section presents a comparative analysis of MAPPO, MADQN, and MADDPG in terms of latency, SINR, throughput, energy efficiency, reward behavior, drone movement patterns, and computational complexity, across both urban and rural scenarios.

\subsubsection{Latency Analysis in Urban and Rural Environments}

We first compare the latency performance of the three algorithms in the two environments.

\paragraph{Urban environment}

\begin{figure}[htbp]
    \footnotesize
    \centering
    % Subplots for Urban Environment
    \subfloat[\footnotesize MADDPG latency in urban environment]{%
        \includegraphics[width=0.4\textwidth]{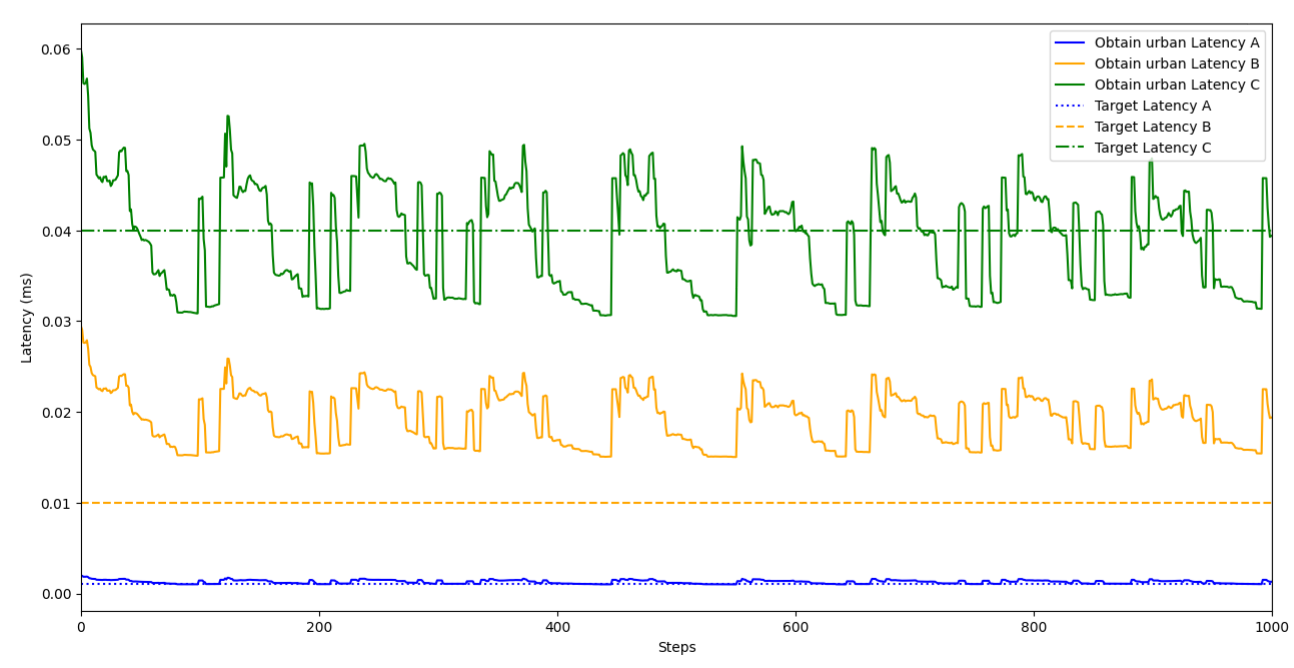}
        \label{fig:latency_ddpg}
    }
    \hfill
    \subfloat[\footnotesize MADQN latency in urban environment]{%
        \includegraphics[width=0.4\textwidth]{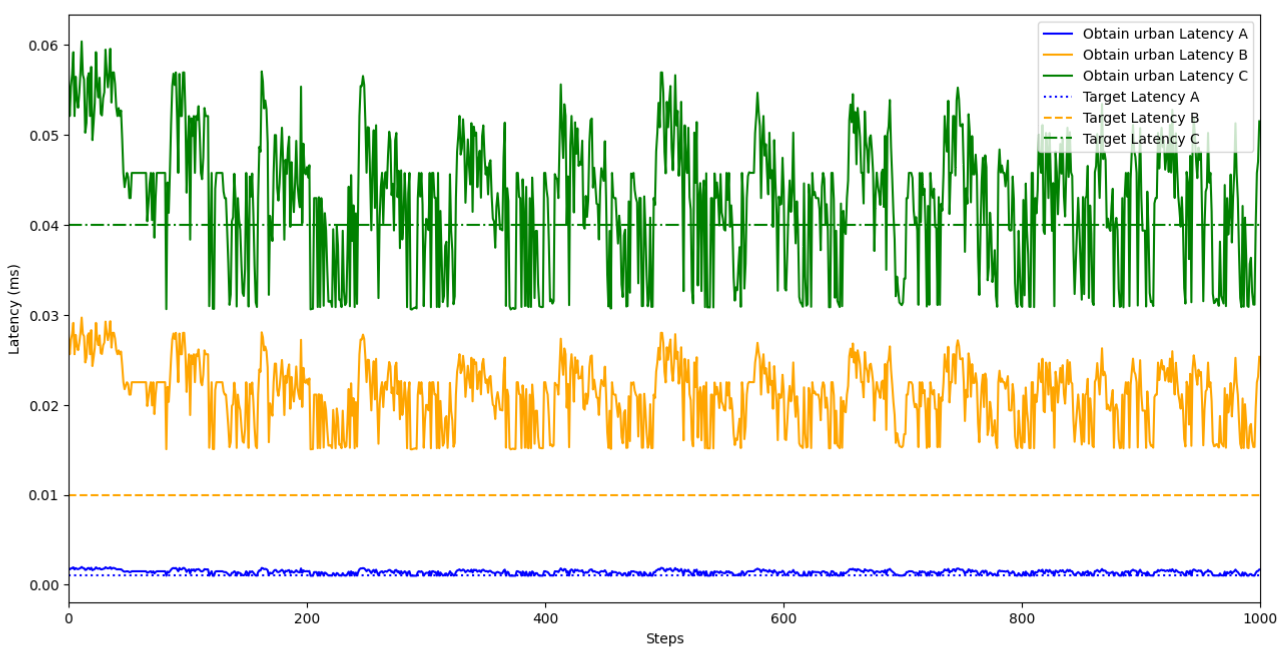}
        \label{fig:latency_DQN}
    }
    \hfill
    \subfloat[\footnotesize MAPPO latency in urban environment]{%
        \includegraphics[width=0.4\textwidth]{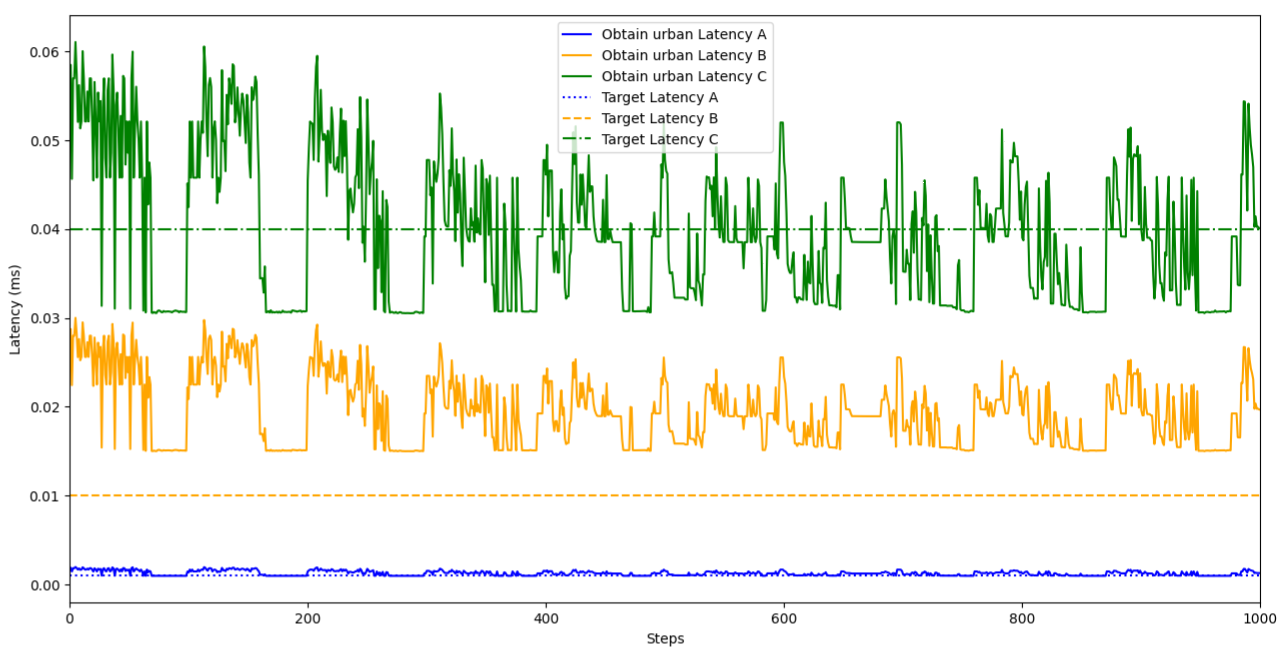}
        \label{fig:latency_PPO}
    }
    \caption{\footnotesize Latency performance in urban environments for MADDPG, MADQN, and MAPPO.}
    \label{fig:latency_urban_combined}
\end{figure}

Fig.~\ref{fig:latency_urban_combined} reports latency in the urban scenario. MAPPO exhibits the best latency performance, consistently maintaining latency below $0.01$~ms for all user tiers, as shown in Fig.~\ref{fig:latency_PPO}. This highlights its strong real-time responsiveness in highly dynamic, interference-heavy conditions. MADQN (Fig.~\ref{fig:latency_DQN}) achieves moderate performance, with latency typically fluctuating between $0.015$~ms and $0.025$~ms. Although relatively stable, it does not reliably meet the most stringent latency targets. MADDPG (Fig.~\ref{fig:latency_ddpg}) records the highest latency, with values occasionally reaching $0.06$~ms and exhibiting irregular patterns, indicating that it adapts less effectively to rapid changes in the urban environment.

\paragraph{Rural environment}

\begin{figure}[htbp]
    \footnotesize
    \centering
    % Subplots for Rural Environment
    \subfloat[\footnotesize MADDPG latency in rural environment]{%
        \includegraphics[width=0.4\textwidth]{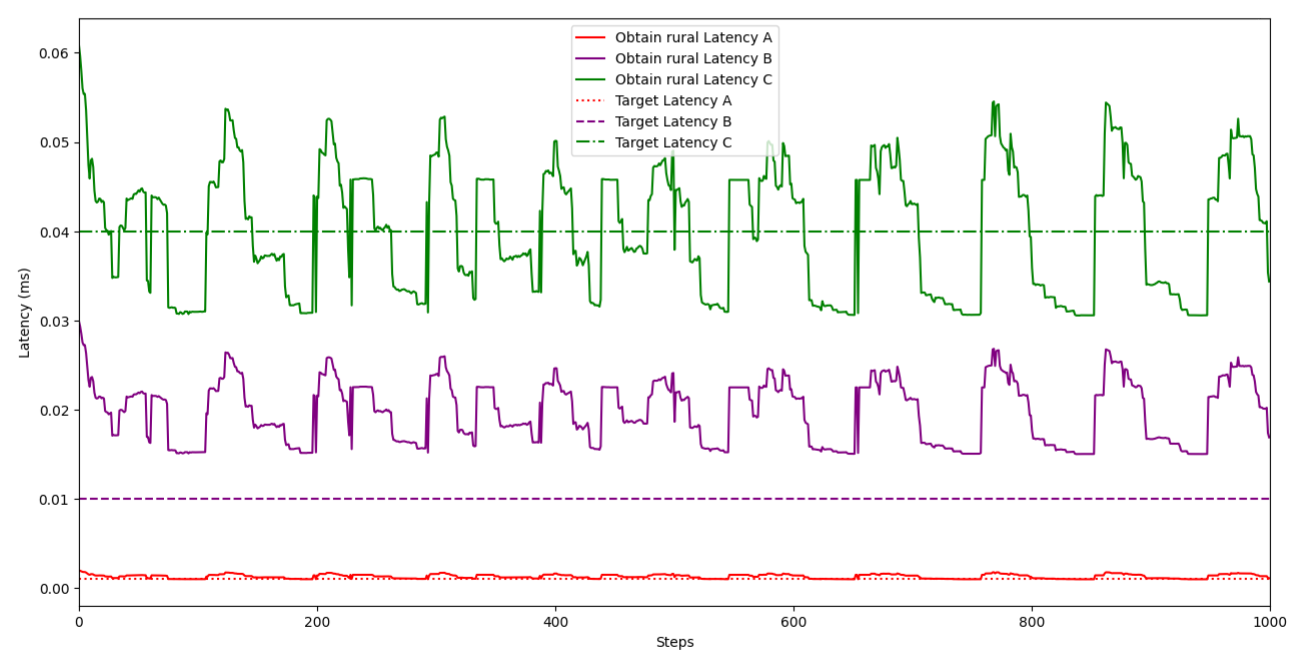}
        \label{fig:latency_ddpg_rular}
    }
    \hfill
    \subfloat[\footnotesize MADQN latency in rural environment]{%
        \includegraphics[width=0.4\textwidth]{Rular/DDPG_rular_1/Latency_rular2_2.PNG} % check path for DQN
        \label{fig:latency_DQN_rular}
    }
    \hfill
    \subfloat[\footnotesize MAPPO latency in rural environment]{%
        \includegraphics[width=0.4\textwidth]{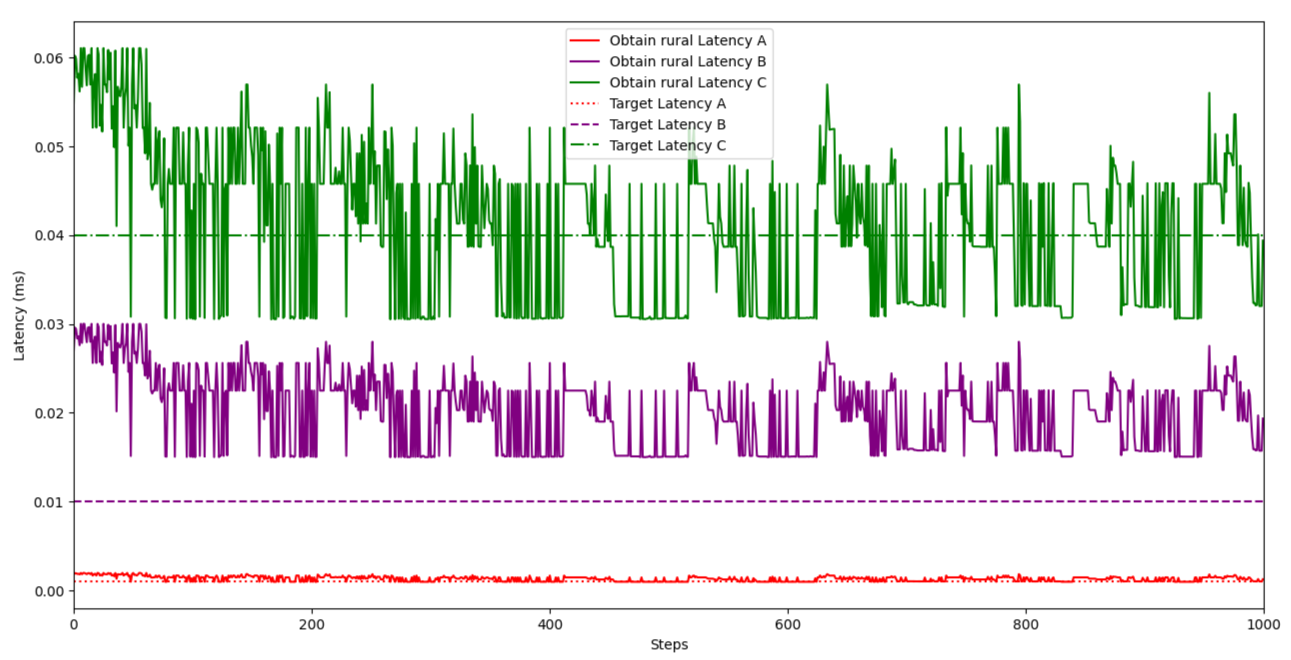}
        \label{fig:latency_PPO_rular}
    }
    \caption{\footnotesize Latency performance (in milliseconds) in rural environments for MADDPG, MADQN, and MAPPO.}
    \label{fig:latency_rural_combined}
\end{figure}

Latency performance in the rural scenario is summarized in Fig.~\ref{fig:latency_rural_combined}. MAPPO (Fig.~\ref{fig:latency_PPO_rular}) again maintains latency below $0.01$~ms, mirroring its behavior in the urban setting and confirming its robustness under less dynamic but wide-area conditions. MADQN (Fig.~\ref{fig:latency_DQN_rular}) shows similar behavior to the urban case, with latency between $0.015$~ms and $0.025$~ms and occasional difficulty in meeting the $0.01$~ms target. MADDPG (Fig.~\ref{fig:latency_ddpg_rular}) exhibits the highest latency and largest fluctuations, reaching up to $0.06$~ms.

Overall, MAPPO emerges as the most effective algorithm for latency-sensitive services in both urban and rural environments. MADQN provides acceptable but not optimal performance, while MADDPG is the least suitable for applications with stringent latency requirements.

\subsubsection{SINR Analysis}

We now compare SINR performance for the three algorithms.

\paragraph{Rural environment}

\begin{figure}[htbp]
    \centering
    \footnotesize
    % Rural SINR Subplots
    \subfloat[\footnotesize MAPPO SINR in rural environment]{%
        \includegraphics[width=0.4\textwidth]{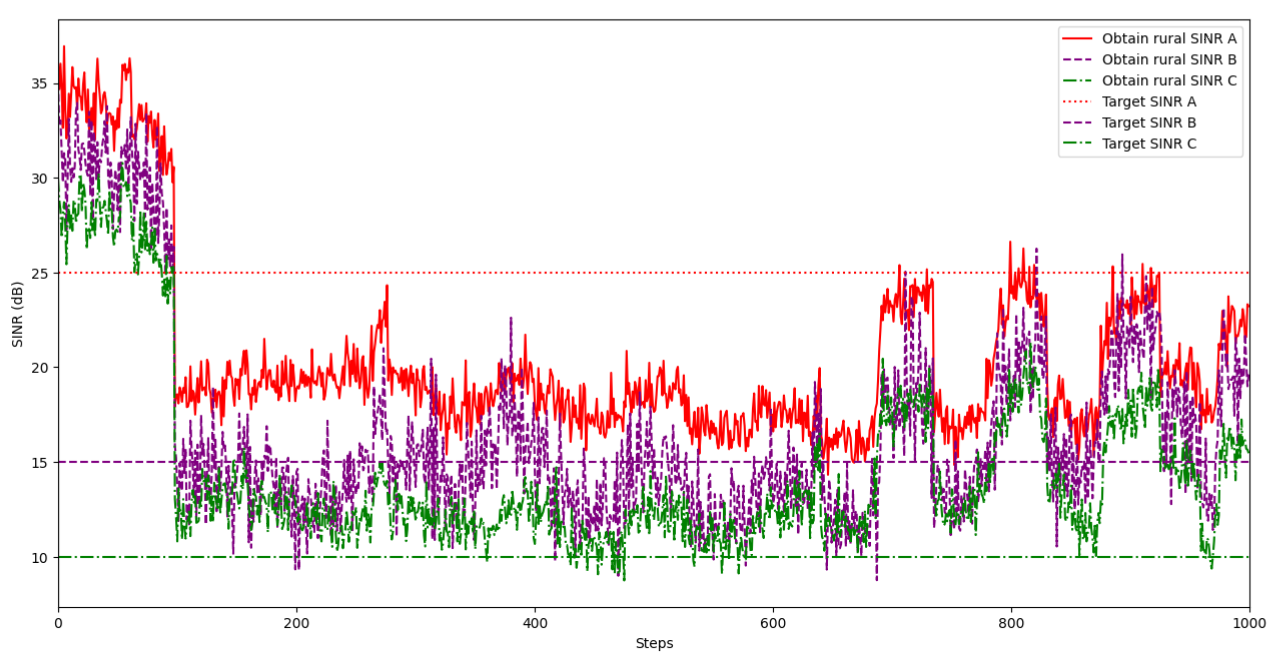}
        \label{fig:sinr_ppo_rular}
    }
    \hfill
    \subfloat[\footnotesize MADQN SINR in rural environment]{%
        \includegraphics[width=0.4\textwidth]{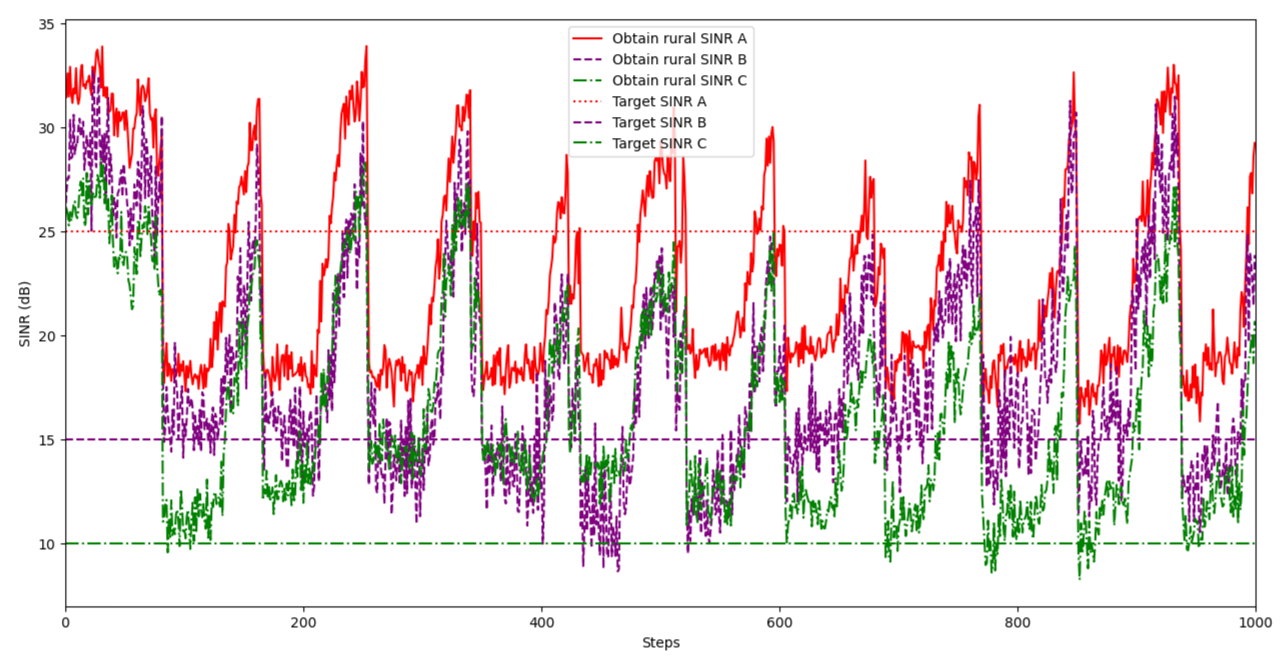}
        \label{fig:sinr_dqn_rular}
    }
    \hfill
    \subfloat[\footnotesize MADDPG SINR in rural environment]{%
        \includegraphics[width=0.4\textwidth]{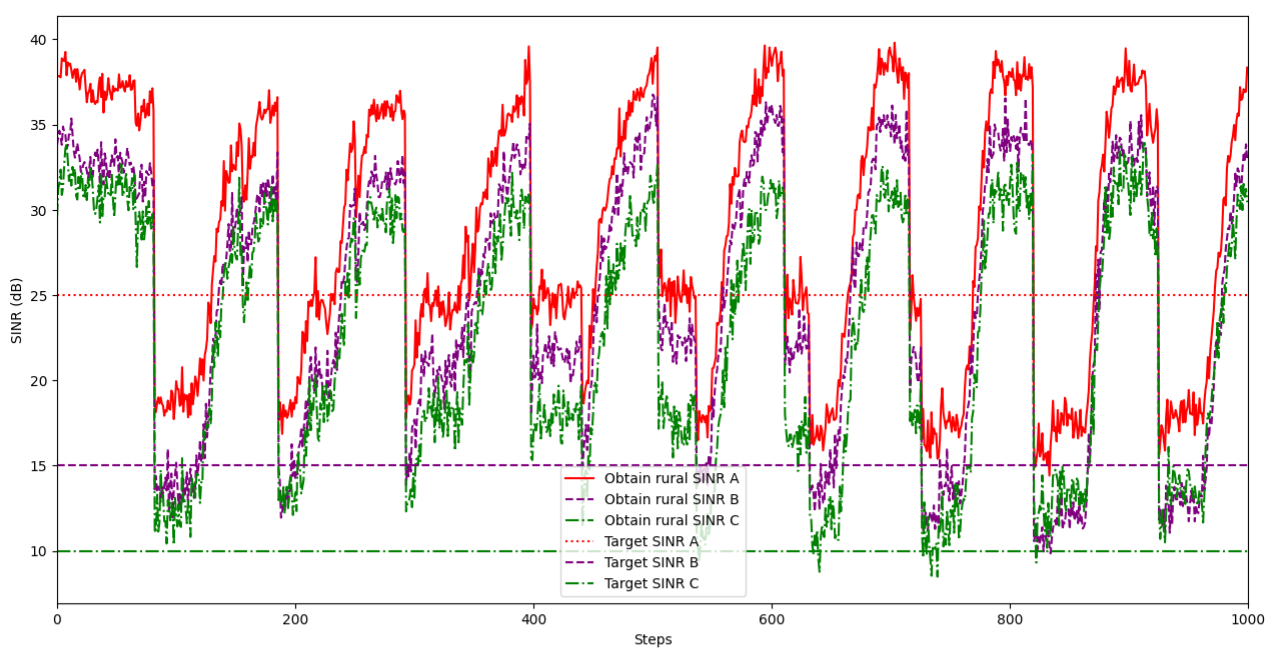}
        \label{fig:sinr_ddpg_rular}
    }
   \caption{\footnotesize SINR performance (in dB) in rural environments. The dashed line marks the 25~dB target for Tier-A users.}
    \label{fig:sinr_rural}
\end{figure}

In the rural scenario (Fig.~\ref{fig:sinr_rural}), MAPPO (Fig.~\ref{fig:sinr_ppo_rular}) consistently achieves SINR levels above the $25$~dB target for Tier-A users, with peaks around $35$~dB, indicating strong interference management in open areas. MADQN (Fig.~\ref{fig:sinr_dqn_rular}) fluctuates between $15$~dB and $25$~dB and often falls below the target, revealing limited adaptability to varying interference conditions. MADDPG (Fig.~\ref{fig:sinr_ddpg_rular}) shows highly variable SINR between $10$~dB and $30$~dB, reflecting instability and weaker interference handling.

\paragraph{Urban environment}

\begin{figure}[htbp]
    \centering
    \footnotesize
    % Urban SINR Subplots
    \subfloat[\footnotesize MAPPO SINR in urban environment]{%
        \includegraphics[width=0.4\textwidth]{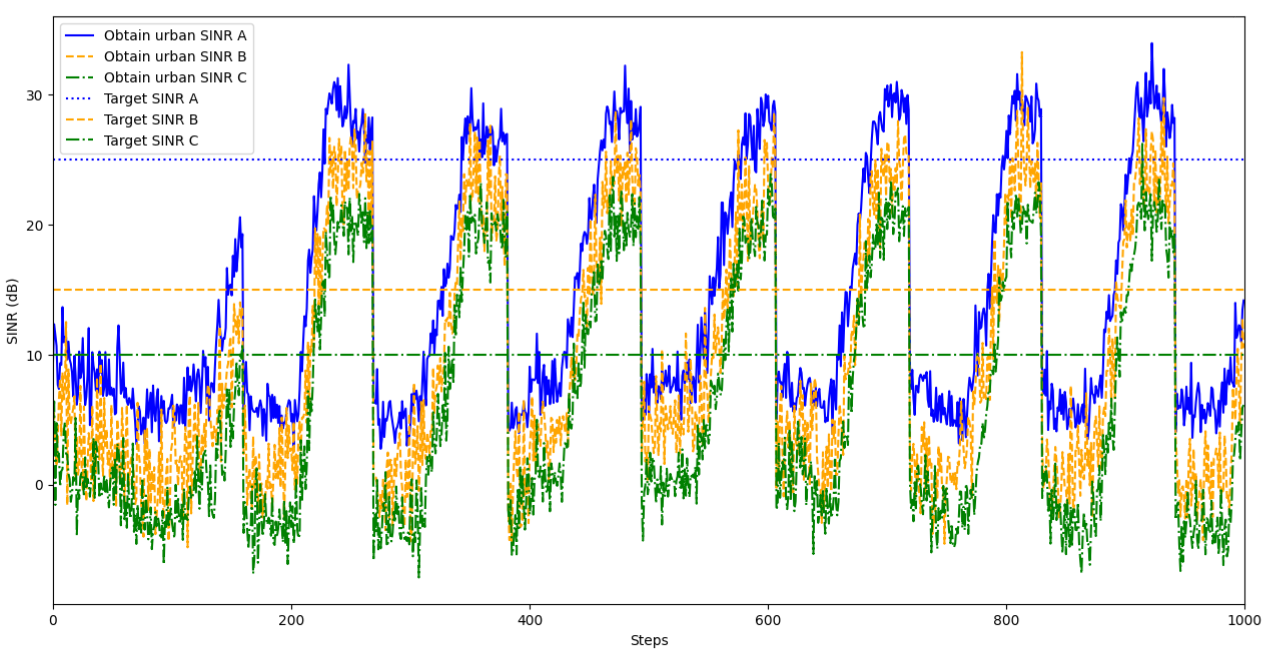}
        \label{fig:sinr_ppo}
    }
    \hfill
    \subfloat[\footnotesize MADQN SINR in urban environment]{%
        \includegraphics[width=0.4\textwidth]{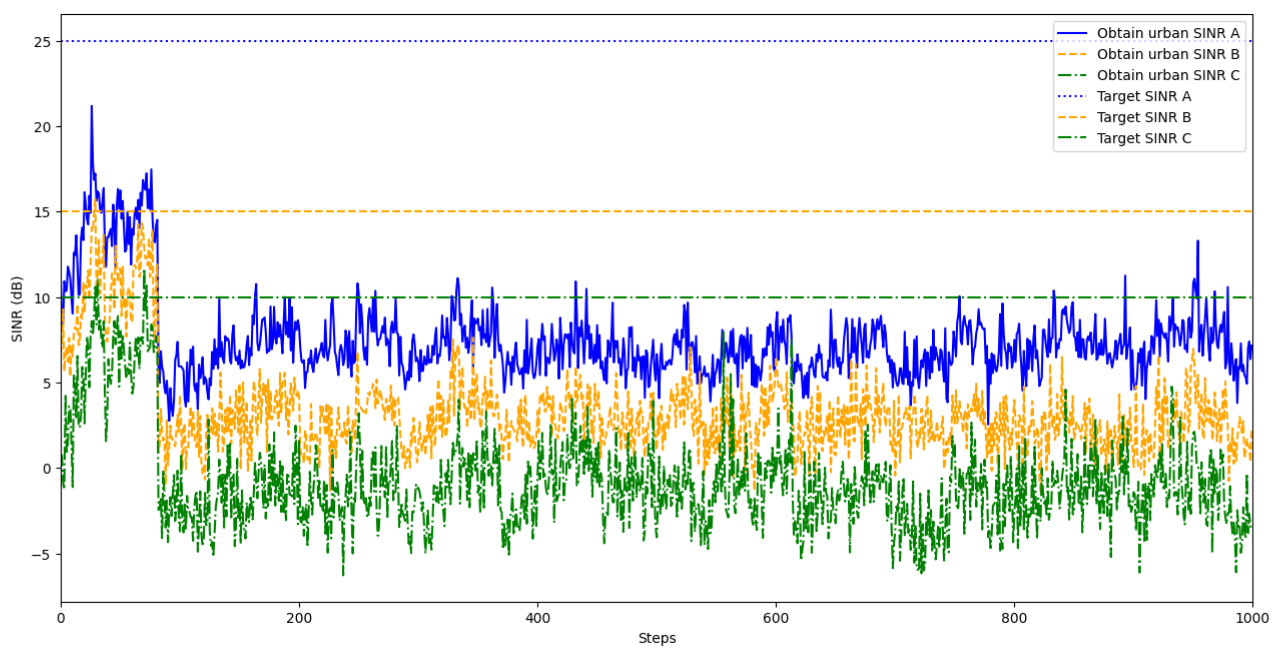}
        \label{fig:sinr_dqn}
    }
    \hfill
    \subfloat[\footnotesize MADDPG SINR in urban environment]{%
        \includegraphics[width=0.4\textwidth]{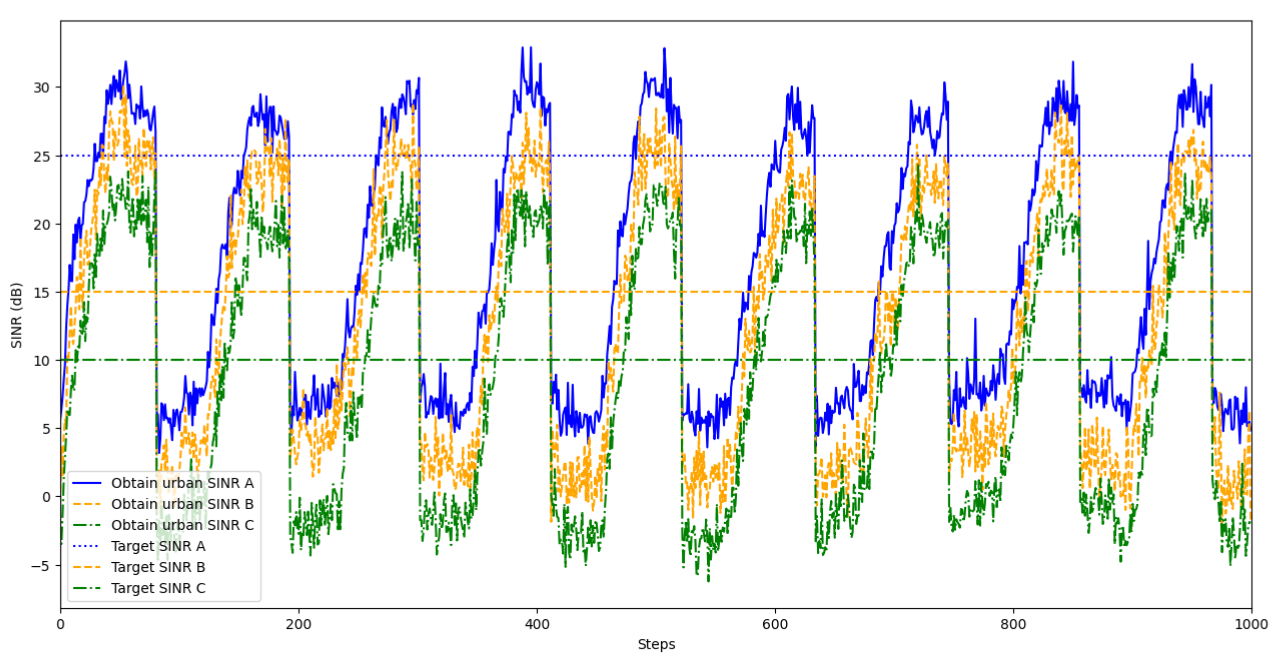}
        \label{fig:sinr_ddpg}
    }
    \caption{\footnotesize SINR performance in urban environments for MAPPO, MADQN, and MADDPG.}
    \label{fig:sinr_urban}
\end{figure}

In the urban environment (Fig.~\ref{fig:sinr_urban}), MAPPO (Fig.~\ref{fig:sinr_ppo}) again delivers the best SINR, most of the time above $25$~dB and peaking near $30$~dB. MADQN (Fig.~\ref{fig:sinr_dqn}) remains in the $10$--$20$~dB range, occasionally meeting a $15$~dB target but failing to consistently ensure high SINR. MADDPG (Fig.~\ref{fig:sinr_ddpg}) performs worst, often remaining between $5$~dB and $20$~dB and frequently below acceptable thresholds.

In summary, MAPPO is the most reliable algorithm from an SINR perspective in both environments, while MADQN and MADDPG provide only moderate or unstable performance.

\subsubsection{Throughput Analysis in Urban and Rural Environments}

We next analyze throughput performance for the three algorithms.

\paragraph{Urban environment}

\begin{figure}[htbp]
    \centering
    \footnotesize
    % Urban Throughput Subplots
    \subfloat[\footnotesize MAPPO throughput in urban environment]{%
        \includegraphics[width=0.45\textwidth]{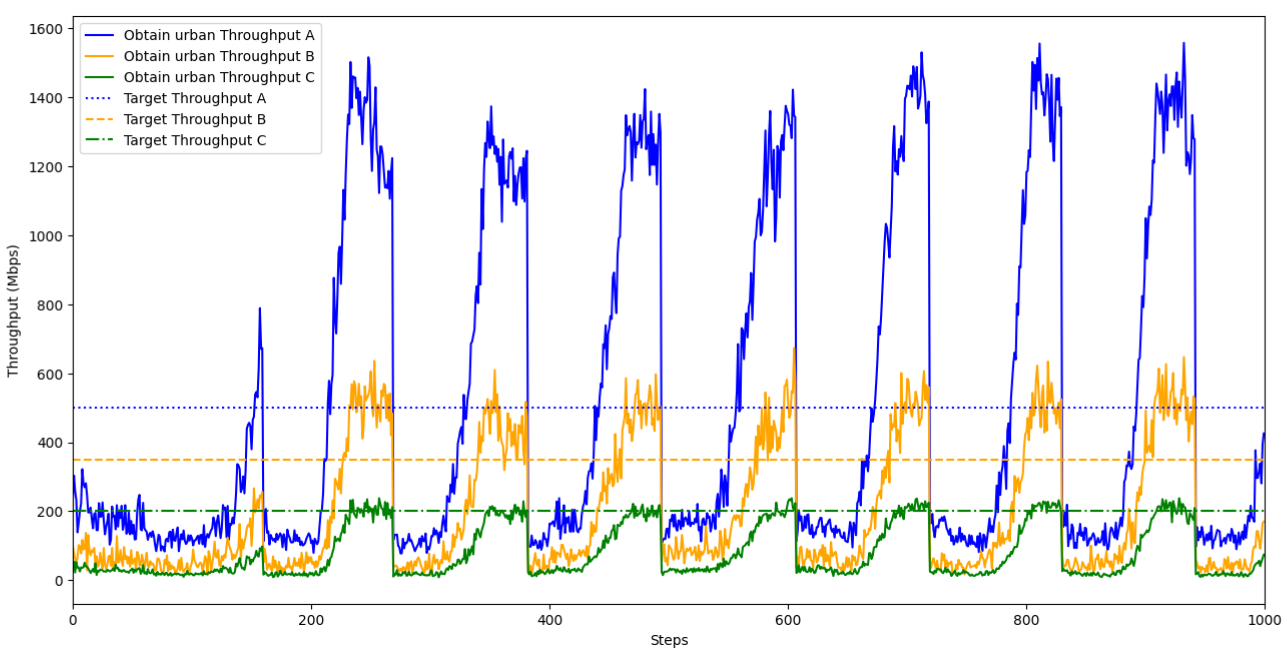}
        \label{fig:throughput_ppo}
    }
    \hfill
    \subfloat[\footnotesize MADQN throughput in urban environment]{%
        \includegraphics[width=0.45\textwidth]{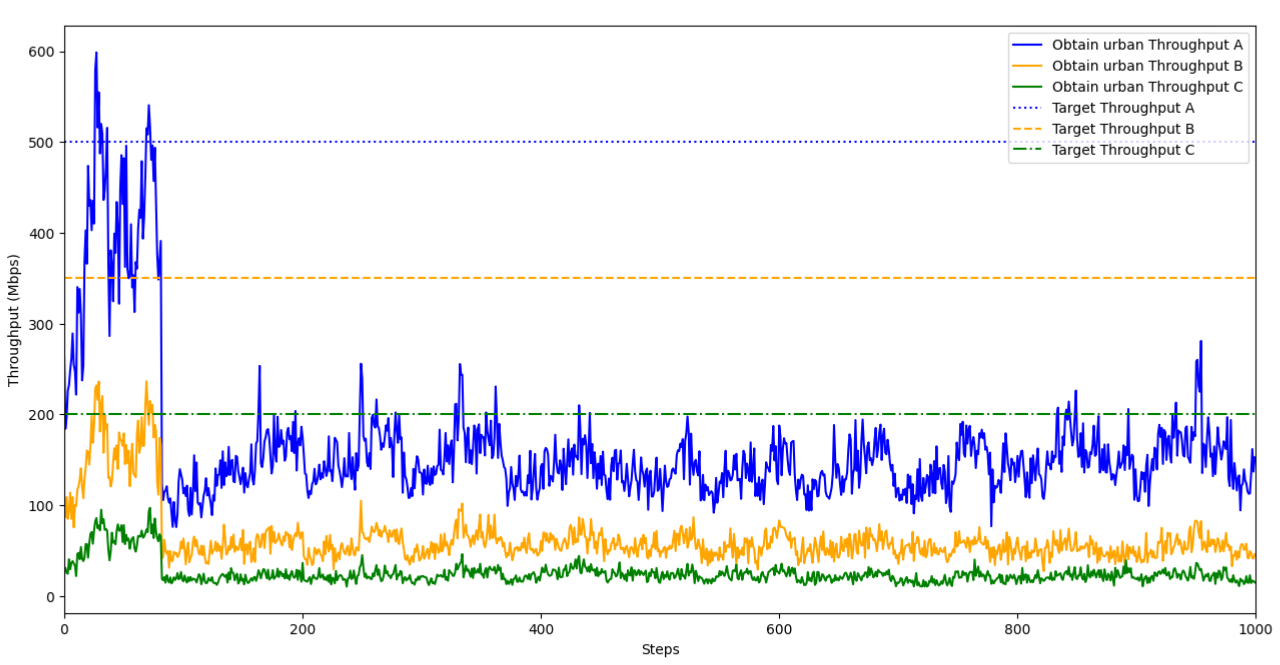}
        \label{fig:throughput_dqn}
    }
    \hfill
    \subfloat[\footnotesize MADDPG throughput in urban environment]{%
        \includegraphics[width=0.45\textwidth]{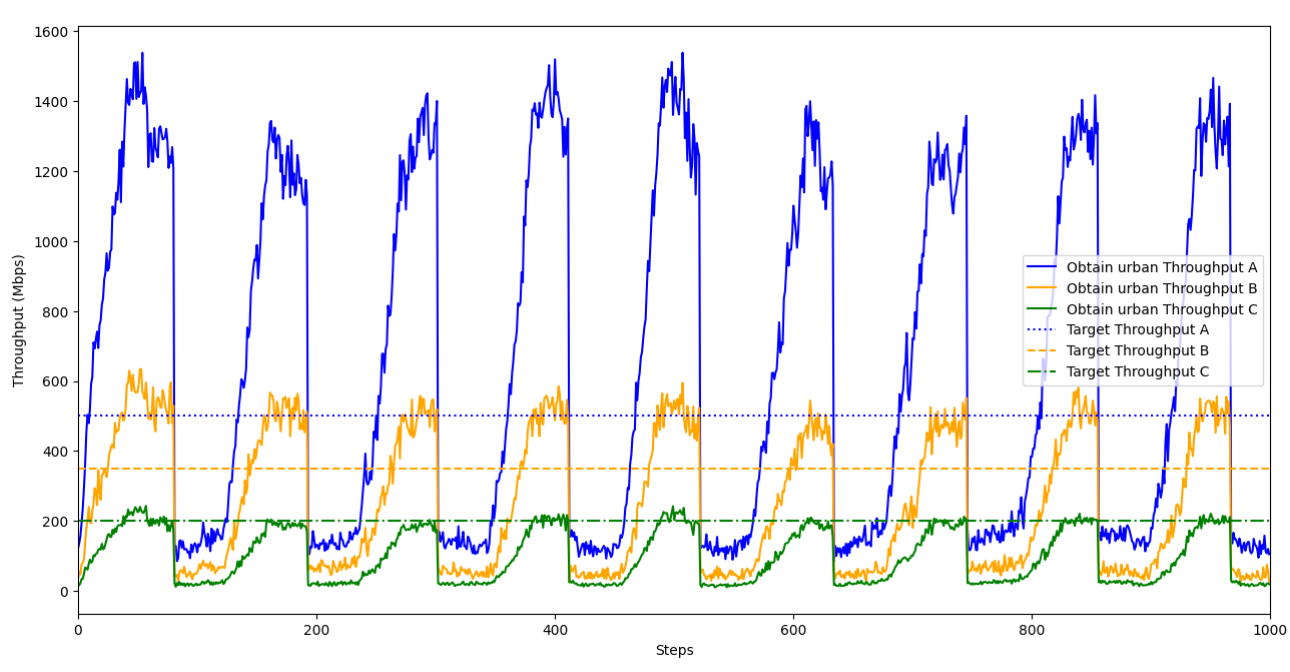}
        \label{fig:throughput_ddpg}
    }
    \caption{\footnotesize Throughput performance in urban environments for MAPPO, MADQN, and MADDPG. Peak values above target thresholds correspond to temporary clustering of UAVs over dense user regions.}
    \label{fig:throughput_urban}
\end{figure}

Fig.~\ref{fig:throughput_urban} compares throughput in the urban scenario. MAPPO (Fig.~\ref{fig:throughput_ppo}) consistently achieves the highest throughput, with Tier-A throughput peaking at approximately $1600$~Mbps and remaining well above the $500$~Mbps target. MADQN (Fig.~\ref{fig:throughput_dqn}) reaches around $600$~Mbps for Tier-A, but struggles to meet target levels for Tiers B and C, reflecting the limitations of its discrete action space in highly dynamic settings. MADDPG (Fig.~\ref{fig:throughput_ddpg}) reaches peaks near $1600$~Mbps for Tier-A but fails to consistently satisfy the targets for lower-priority services, indicating moderate adaptability in dense urban conditions.

\paragraph{Rural environment}

\begin{figure}[htbp]
    \centering
    \footnotesize
    % Rural Throughput Subplots
    \subfloat[\footnotesize MAPPO throughput in rural environment]{%
        \includegraphics[width=0.45\textwidth]{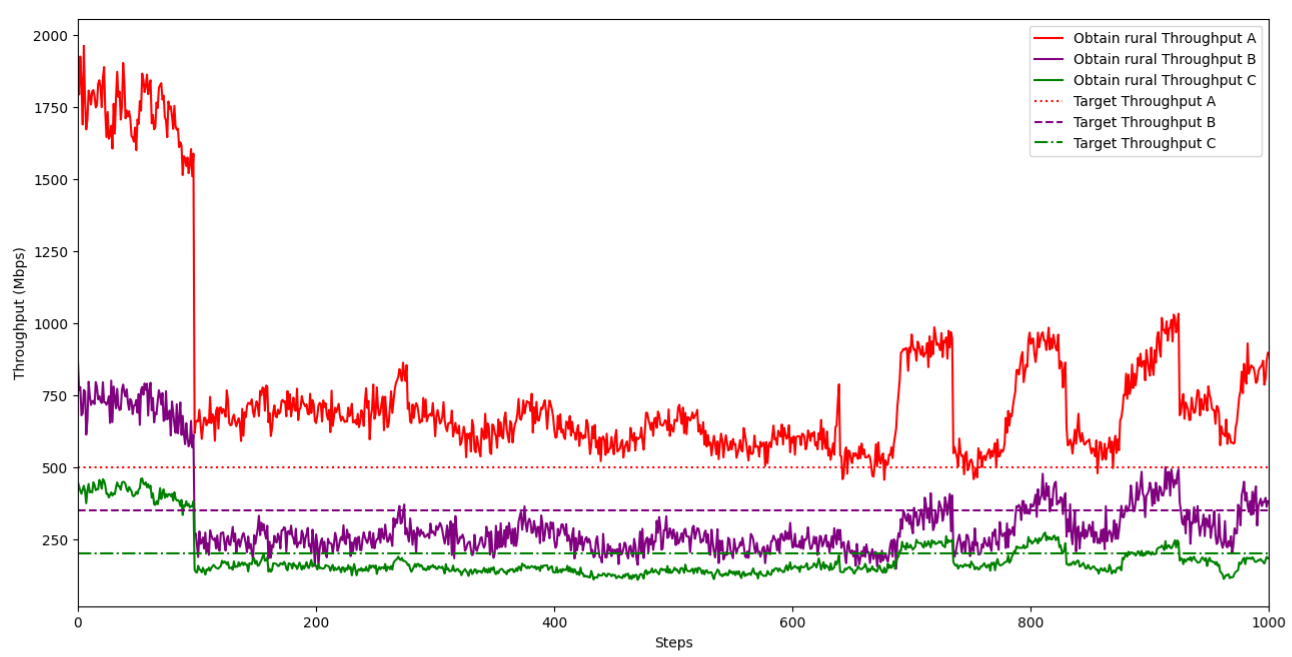}
        \label{fig:throughput_ppo_rular}
    }
    \hfill
    \subfloat[\footnotesize MADQN throughput in rural environment]{%
        \includegraphics[width=0.45\textwidth]{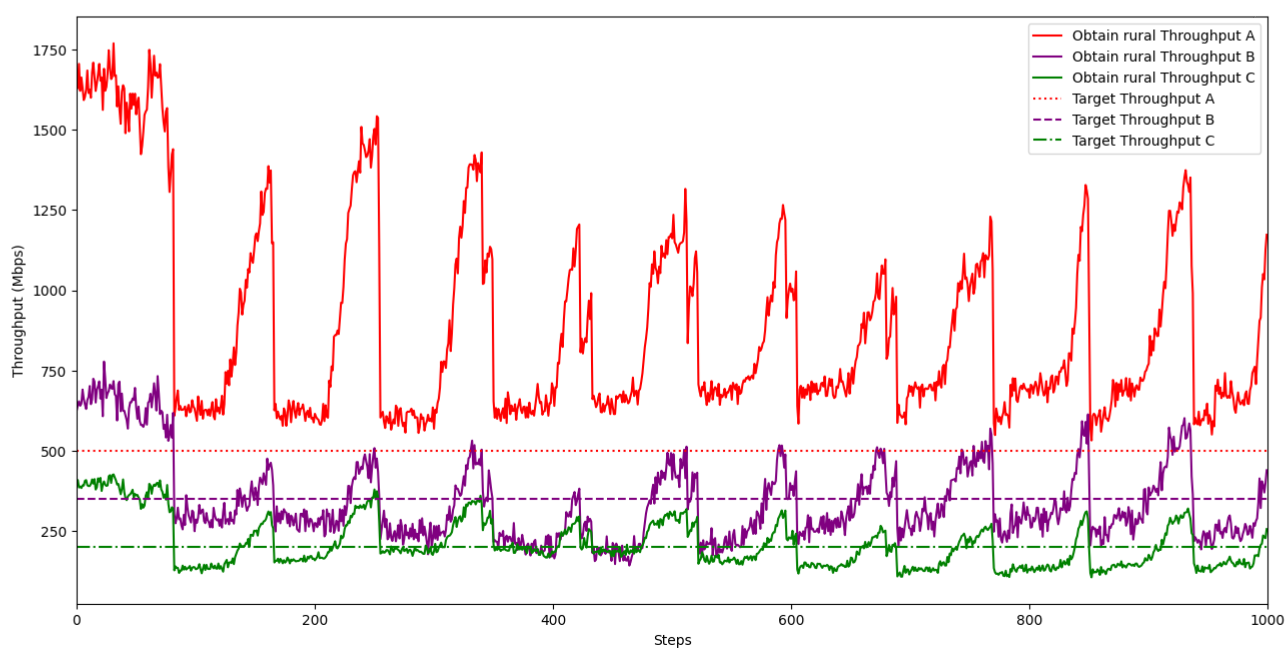}
        \label{fig:throughput_dqn_rular}
    }
    \hfill
    \subfloat[\footnotesize MADDPG throughput in rural environment]{%
        \includegraphics[width=0.45\textwidth]{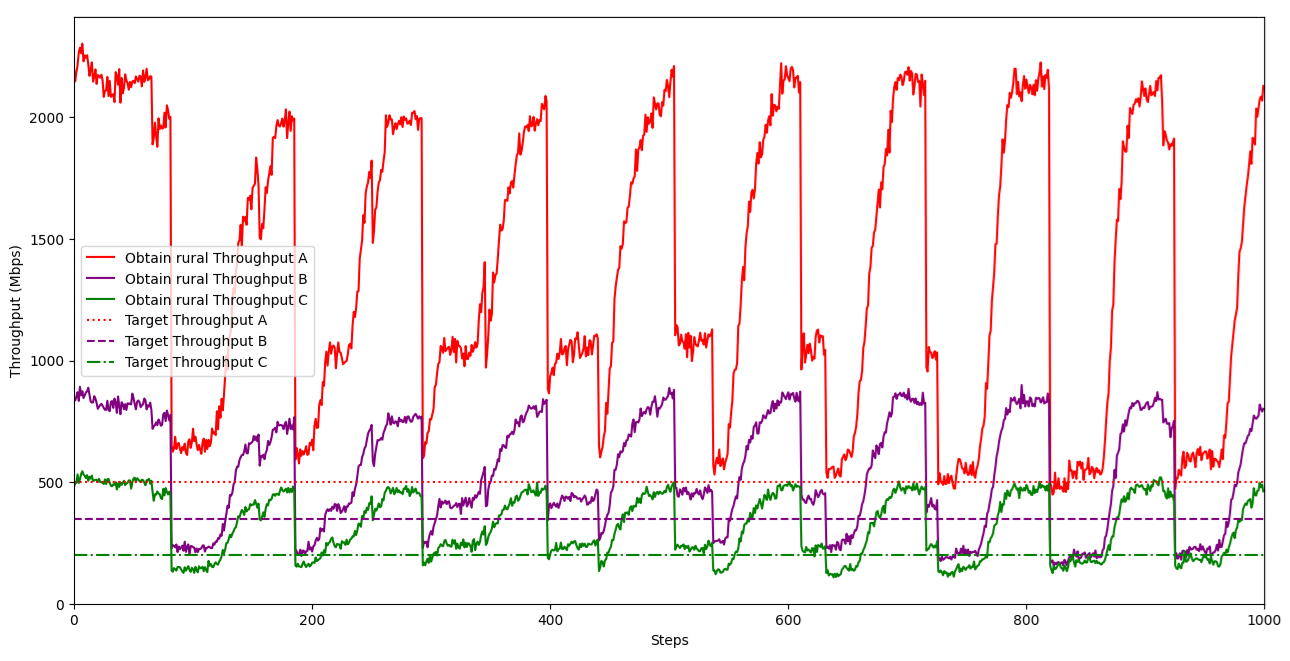}
        \label{fig:throughput_ddpg_rular}
    }
    \caption{\footnotesize Throughput performance in rural environments for MAPPO, MADQN, and MADDPG.}
    \label{fig:throughput_rular}
\end{figure}

In the rural scenario (Fig.~\ref{fig:throughput_rular}), MAPPO (Fig.~\ref{fig:throughput_ppo_rular}) again provides the strongest performance, with Tier-A throughput peaking around $2000$~Mbps and remaining above the $400$~Mbps target for most of the time. MADQN (Fig.~\ref{fig:throughput_dqn_rular}) shows periodic spikes for Tier-A but has difficulty sustaining high throughput. MADDPG (Fig.~\ref{fig:throughput_ddpg_rular}) achieves Tier-A throughput up to about $1750$~Mbps and provides somewhat more stable throughput for Tiers B and C than MADQN, but still does not consistently meet all targets.

Overall, MAPPO delivers the best throughput across both environments, especially under high-demand conditions.

\subsubsection{Energy Efficiency in Urban and Rural Environments}

Energy management is critical for UAVs with limited battery capacity. We analyze battery level trajectories in both environments.

\paragraph{Rural environment}

\begin{figure}[!htbp]
    \centering
    \footnotesize
    % Rural Subplots
    \subfloat[\footnotesize MADDPG rural battery levels]{%
        \includegraphics[width=0.4\textwidth]{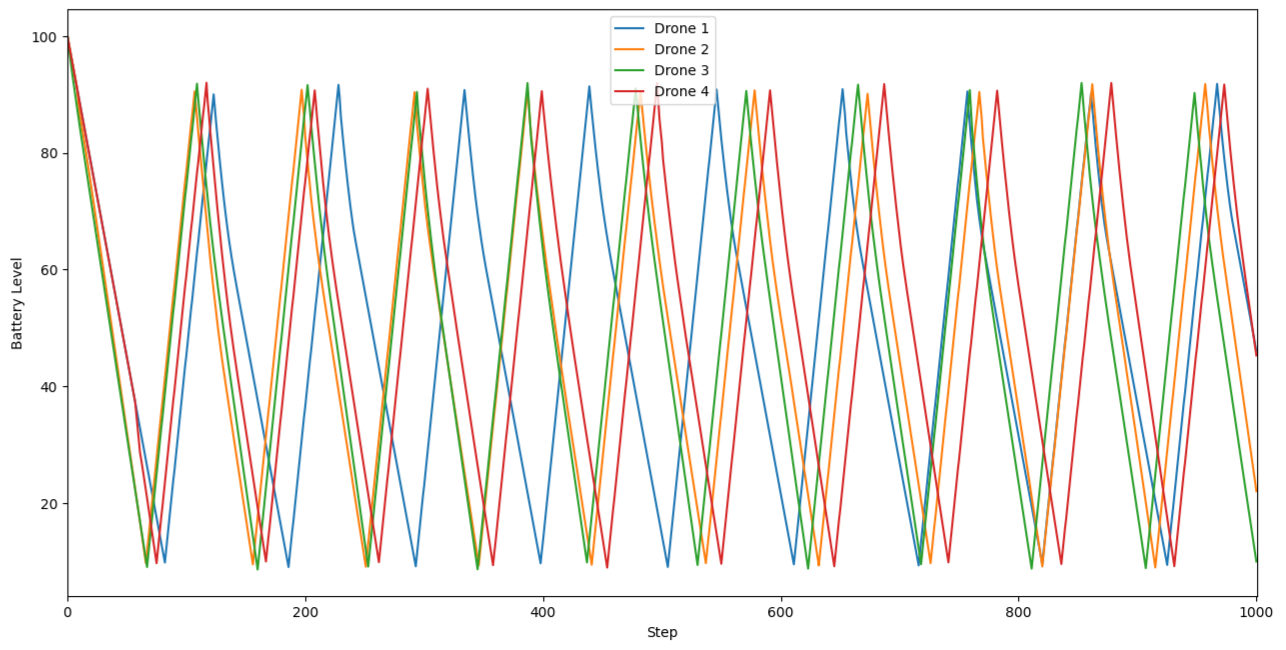}
        \label{fig:rural_battery_ddpg}
    }
    \hfill
    \subfloat[\footnotesize MADQN rural battery levels]{%
        \includegraphics[width=0.4\textwidth]{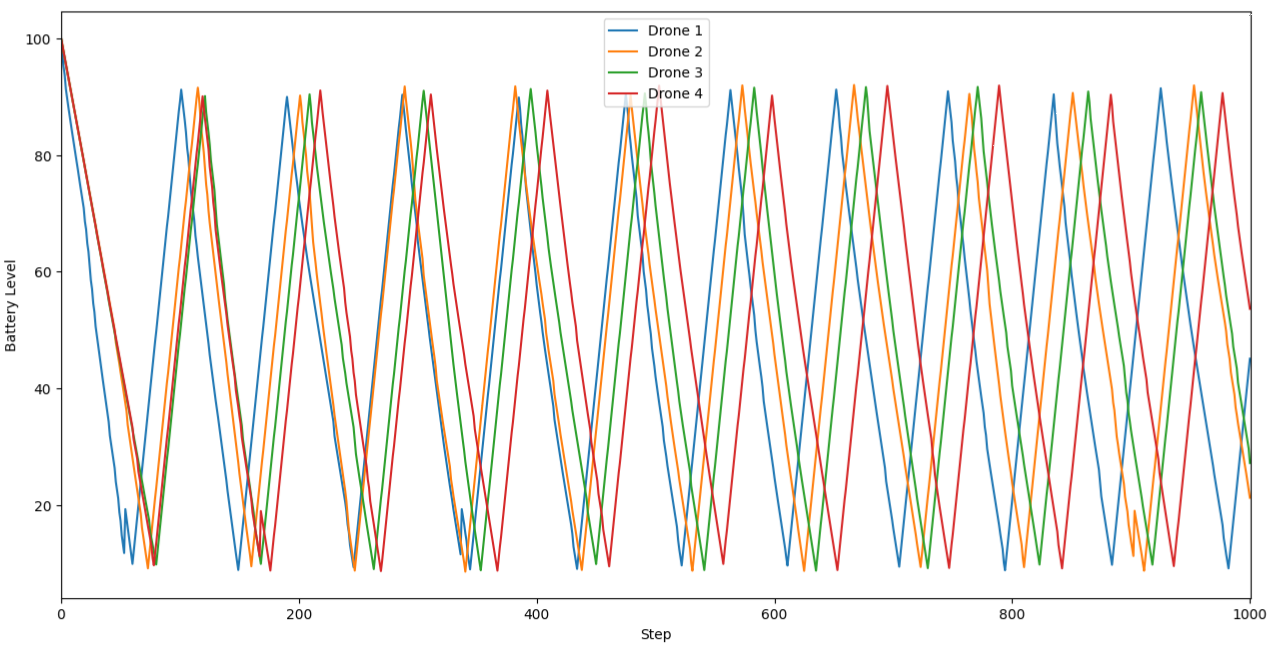}
        \label{fig:rural_battery_dqn}
    }
    \hfill
    \subfloat[\footnotesize MAPPO rural battery levels]{%
        \includegraphics[width=0.4\textwidth]{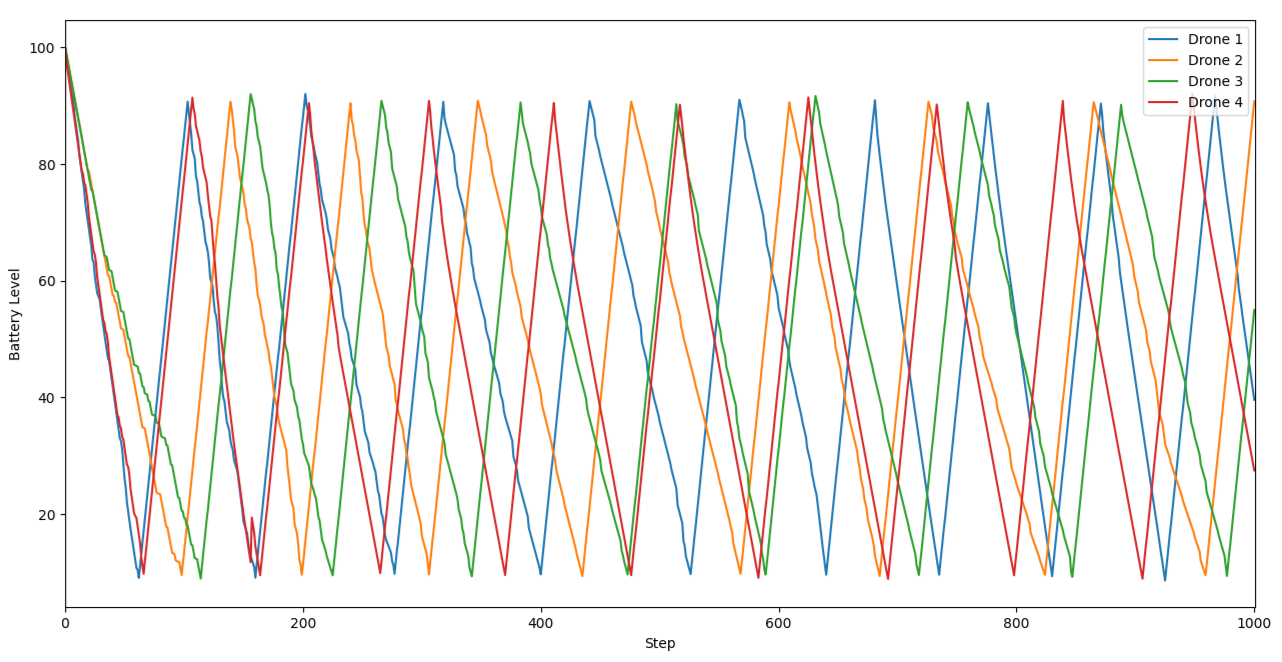}
        \label{fig:rural_battery_ppo}
    }
    \caption{\footnotesize Battery levels in rural environments for MADDPG, MADQN, and MAPPO.}
    \label{fig:battery_rural_combined}
\end{figure}

In rural environments (Fig.~\ref{fig:battery_rural_combined}), UAVs exhibit relatively regular depletion and recharge cycles. MAPPO (Fig.~\ref{fig:rural_battery_ppo}) shows smooth and synchronized battery usage, reflecting efficient policy learning that balances coverage and energy consumption. MADQN (Fig.~\ref{fig:rural_battery_dqn}) behaves similarly, with stable and predictable recharge cycles; the discrete action space is well suited to the slower dynamics of the rural scenario. MADDPG (Fig.~\ref{fig:rural_battery_ddpg}) also maintains acceptable battery levels but with less regular recharge timing, indicating slightly less stable energy management.

\paragraph{Urban environment}

\begin{figure}[!htbp]
    \centering
    \footnotesize
    % Urban Subplots
    \subfloat[\footnotesize MADDPG urban battery levels]{%
        \includegraphics[width=0.4\textwidth]{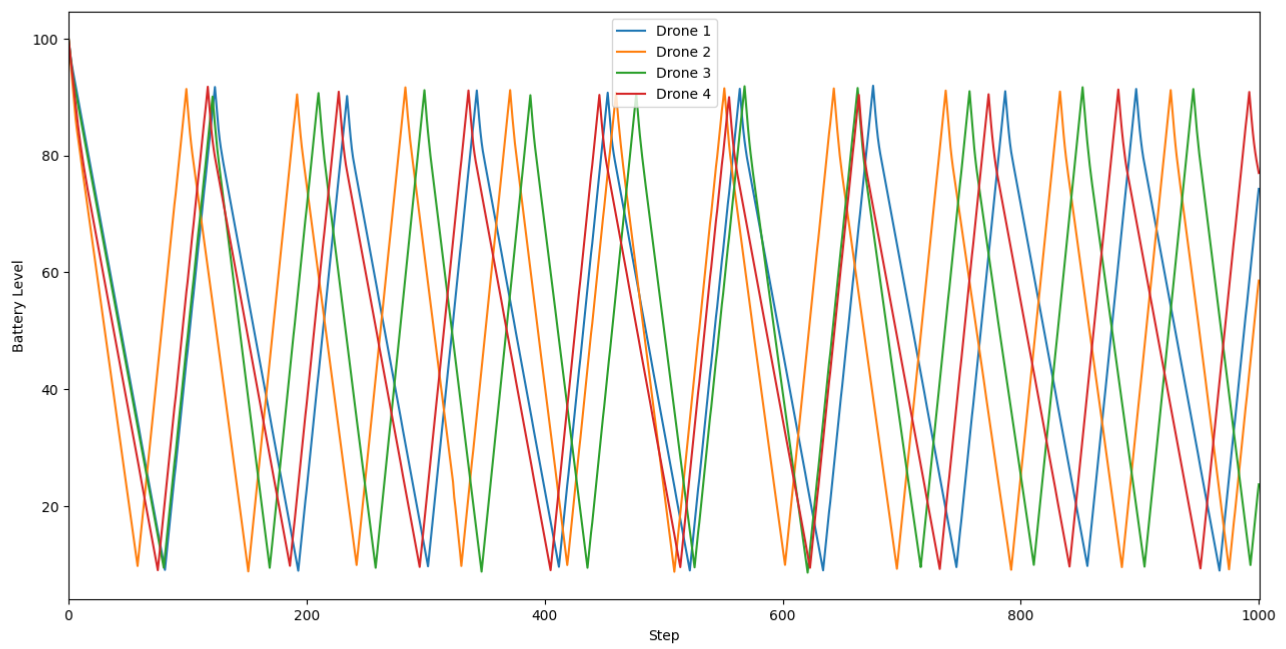}
        \label{fig:urban_battery_ddpg}
    }
    \hfill
    \subfloat[\footnotesize MADQN urban battery levels]{%
        \includegraphics[width=0.4\textwidth]{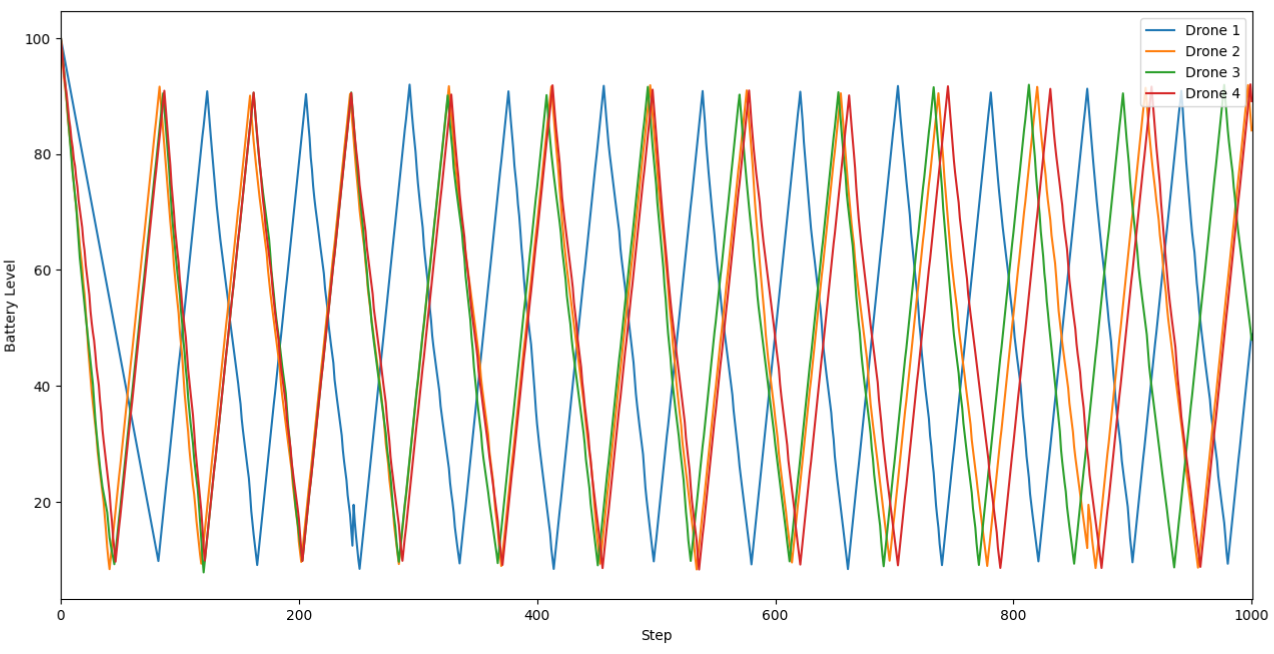}
        \label{fig:urban_battery_dqn}
    }
    \hfill
    \subfloat[\footnotesize MAPPO urban battery levels]{%
        \includegraphics[width=0.4\textwidth]{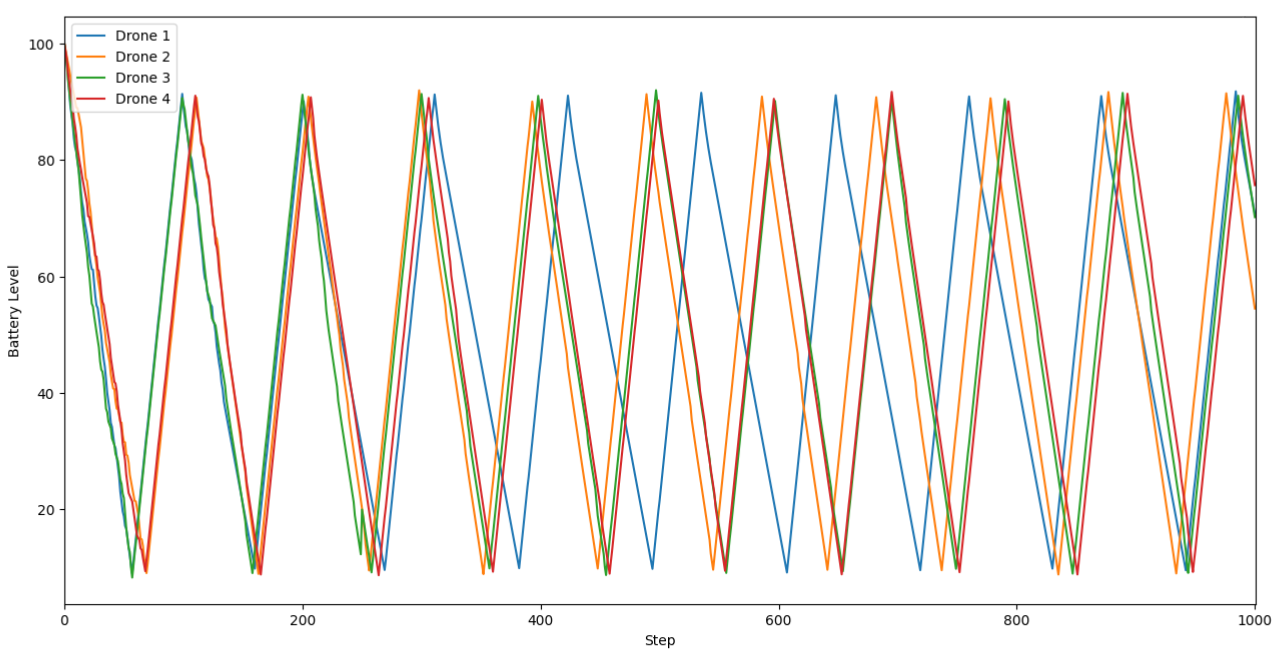}
        \label{fig:urban_battery_ppo}
    }
    \caption{\footnotesize Battery levels in urban environments for MADDPG, MADQN, and MAPPO.}
    \label{fig:battery_urban_combined}
\end{figure}

In urban environments (Fig.~\ref{fig:battery_urban_combined}), energy usage becomes more complex due to frequent repositioning and interference. MAPPO (Fig.~\ref{fig:urban_battery_ppo}) maintains relatively efficient recharge cycles but experiences faster depletion compared to the rural case, as UAVs move more often to serve dense user clusters. MADQN (Fig.~\ref{fig:urban_battery_dqn}) shows balanced but less synchronized energy patterns, requiring more frequent recharges. MADDPG (Fig.~\ref{fig:urban_battery_ddpg}) displays the least efficient behavior, with irregular recharge intervals and frequent low-battery states.

Overall, MAPPO provides the most robust energy management across both environments, MADQN performs reasonably well, and MADDPG is the most variable in energy efficiency.

\subsubsection{Reward analysis for UAV agents}

We now examine the cumulative reward obtained by UAV agents under each algorithm, which reflects the trade-off between QoS and energy consumption.

\paragraph{Rural environment}

In the rural scenario, MADDPG reaches a cumulative reward of approximately $1.0 \times 10^6$, indicating steady but relatively slow improvement. MADQN achieves around $4.5 \times 10^6$, benefitting from its discrete action space to stabilize behavior once an effective policy is found. MAPPO attains approximately $4.0 \times 10^6$, slightly below MADQN but still demonstrating strong performance in balancing QoS and energy under low-interference, wide-area conditions.

\paragraph{Urban environment}

\begin{figure}[htbp]
    \centering
    \includegraphics[width=0.4\textwidth]{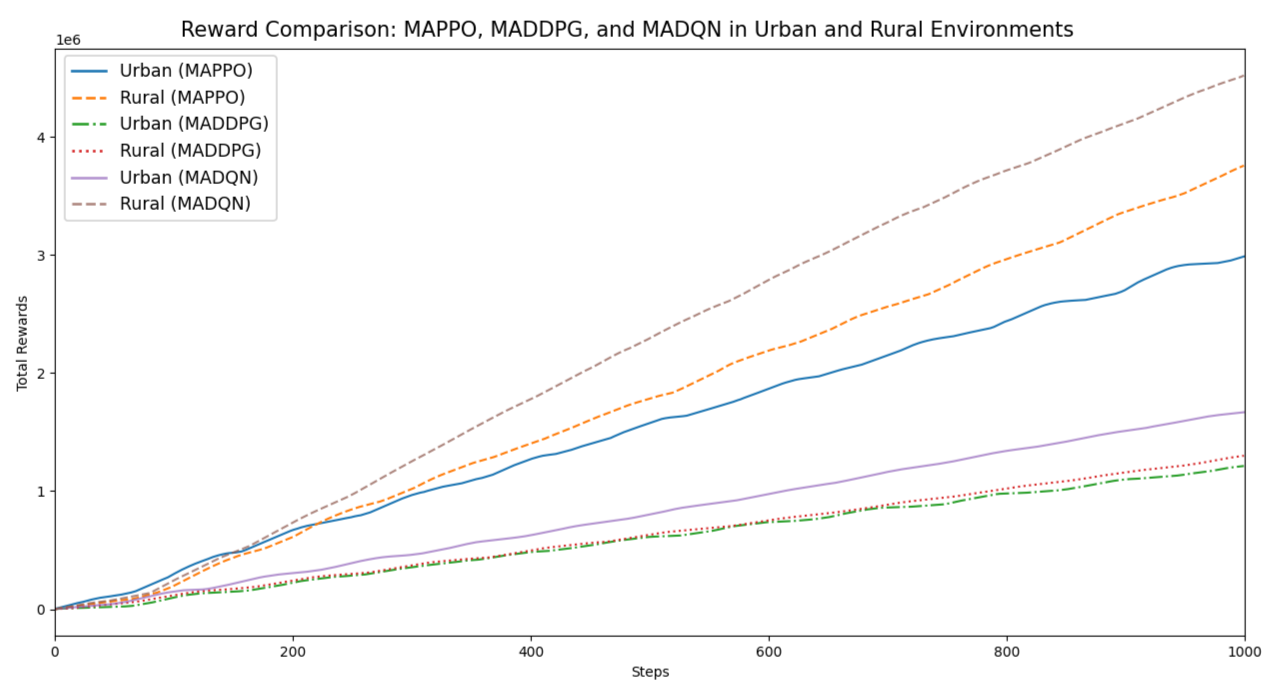}
    \caption{\footnotesize MADDPG cumulative rewards in urban and rural environments.}
    \label{fig:rewards_ddpg}
\end{figure}

In the urban environment (Fig.~\ref{fig:rewards_ddpg}), MADDPG again reaches about $1.0 \times 10^6$ cumulative reward and struggles with the rapid dynamics. MADQN obtains roughly $1.5 \times 10^6$, showing some benefit from discrete control but limited adaptability. MAPPO attains approximately $3.0 \times 10^6$, with a smoother reward curve and better convergence behavior, reflecting its robustness in complex and highly dynamic settings. The reward curves for MADQN and MAPPO follow the same trends as the latency and throughput results and are omitted here for space.

\subsubsection{Drone convergence patterns in urban and rural environments}

We finally analyze the spatial convergence of UAV trajectories for each algorithm.

\paragraph{Urban environment}

\begin{figure}[htbp]
    \centering
    \footnotesize
    % Urban Subplots
    \subfloat[\footnotesize MADDPG urban movements]{%
        \includegraphics[width=0.34\textwidth]{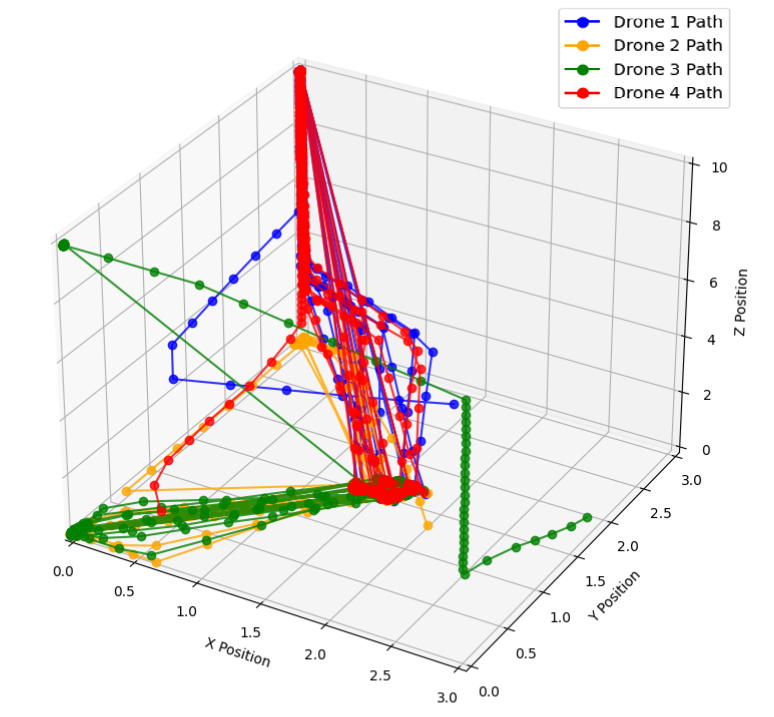}
        \label{fig:urban_movement_ddpg}
    }
    \hfill
    \subfloat[\footnotesize MADQN urban movements]{%
        \includegraphics[width=0.34\textwidth]{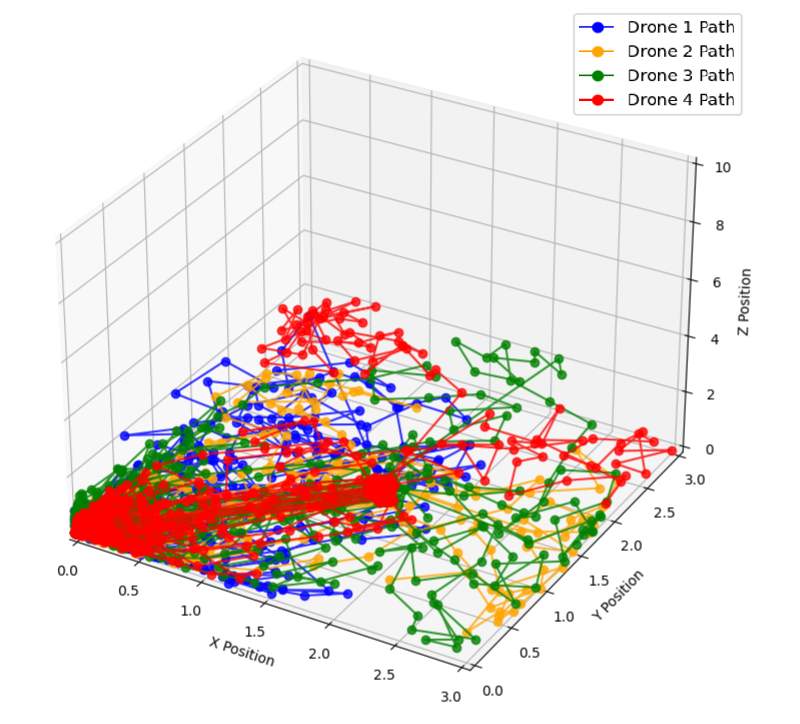}
        \label{fig:urban_movement_dqn}
    }
    \hfill
    \subfloat[\footnotesize MAPPO urban movements]{%
        \includegraphics[width=0.4\textwidth]{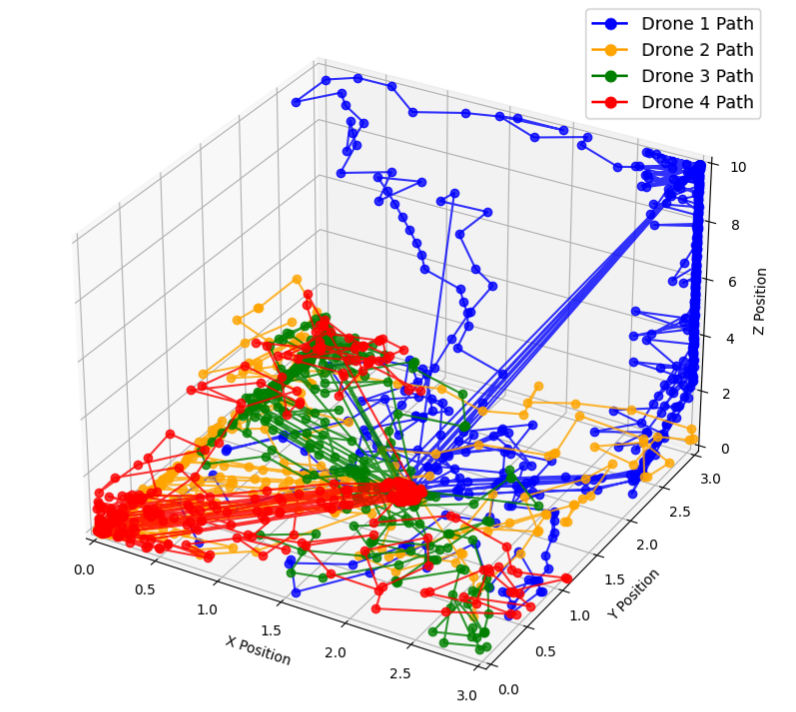}
        \label{fig:urban_movement_ppo}
    }
    \caption{\footnotesize Drone movement patterns in urban environments for MADDPG, MADQN, and MAPPO.}
    \label{fig:drone_movements_comparison}
\end{figure}

In the urban scenario (Fig.~\ref{fig:drone_movements_comparison}), MAPPO (Fig.~\ref{fig:urban_movement_ppo}) produces smooth and well-distributed trajectories, adapting effectively to obstacles and dense user clusters while limiting mutual interference between drones. MADQN (Fig.~\ref{fig:urban_movement_dqn}) leverages its discrete actions to handle obstacles but yields jerkier paths and less consistent altitude control. MADDPG (Fig.~\ref{fig:urban_movement_ddpg}) generates relatively stable but overly conservative paths, prioritizing smoothness over aggressive adaptation, which can reduce coverage efficiency in dense areas.

\paragraph{Rural environment}

\begin{figure}[htbp]
    \centering
    \footnotesize
    % Rural Subplots
    \subfloat[\footnotesize MADDPG rural movements]{%
        \includegraphics[width=0.34\textwidth]{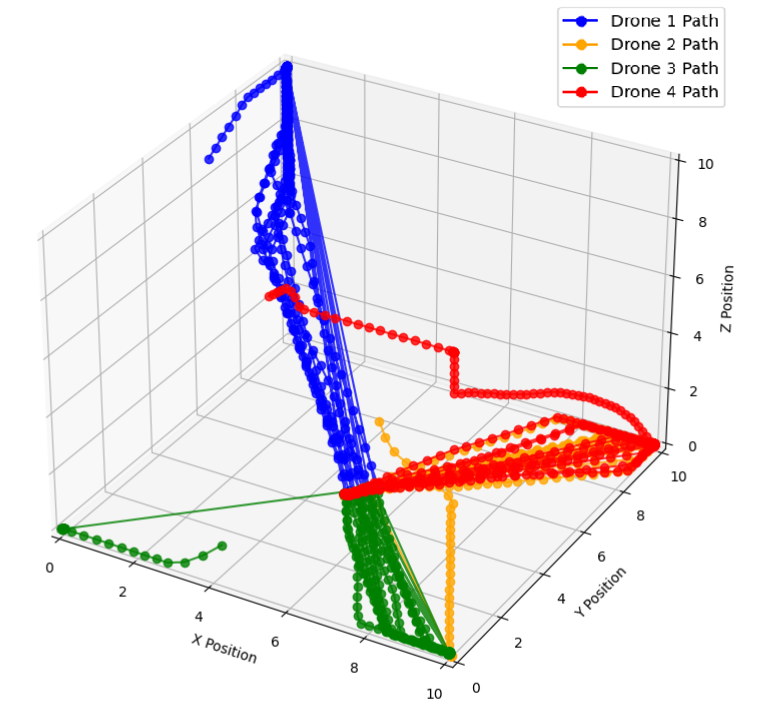}
        \label{fig:rural_movement_ddpg}
    }
    \hfill
    \subfloat[\footnotesize MADQN rural movements]{%
        \includegraphics[width=0.34\textwidth]{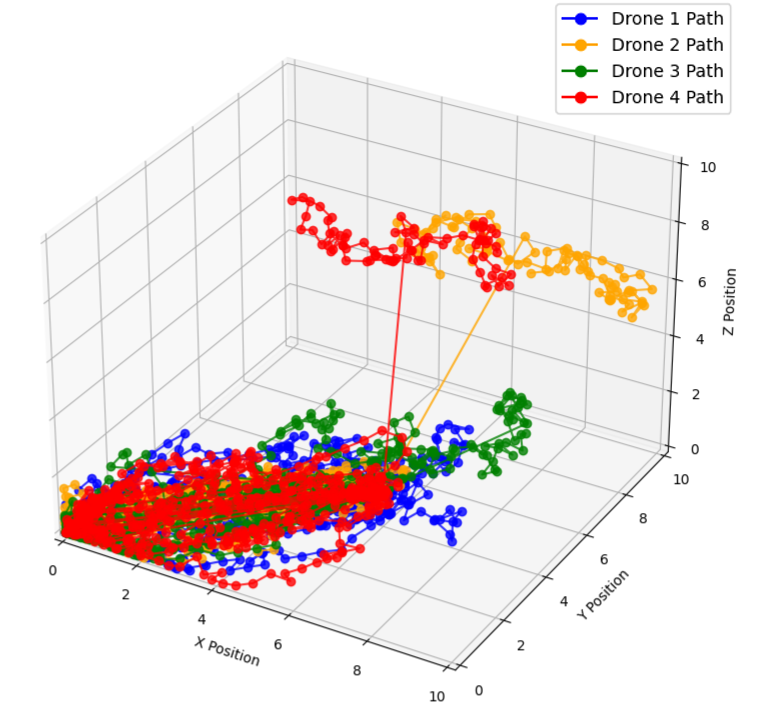}
        \label{fig:rural_movement_dqn}
    }
    \hfill
    \subfloat[\footnotesize MAPPO rural movements]{%
        \includegraphics[width=0.34\textwidth]{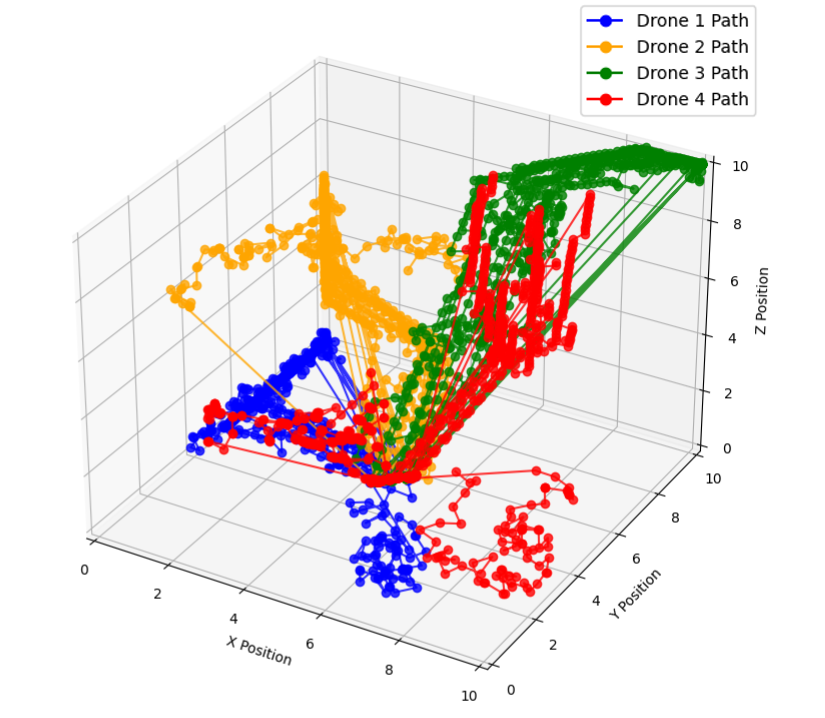}
        \label{fig:rural_movement_ppo}
    }
    \caption{\footnotesize Drone movement patterns in rural environments for MADDPG, MADQN, and MAPPO.}
\end{figure}

In the rural case (Fig.~\ref{fig:rural_movement_ddpg}--\ref{fig:rural_movement_ppo}), MAPPO again produces smooth and energy-efficient trajectories with moderate altitude variations. MADQN disperses drones effectively over the larger area but shows more erratic altitude changes, revealing some inefficiencies in energy usage. MADDPG maintains very stable altitudes and paths, favouring energy conservation over exploration, which is beneficial in low-interference rural conditions but may limit adaptability to sudden user-density changes.

\begin{figure}[htbp]
\centering
\includegraphics[width=0.35\textwidth]{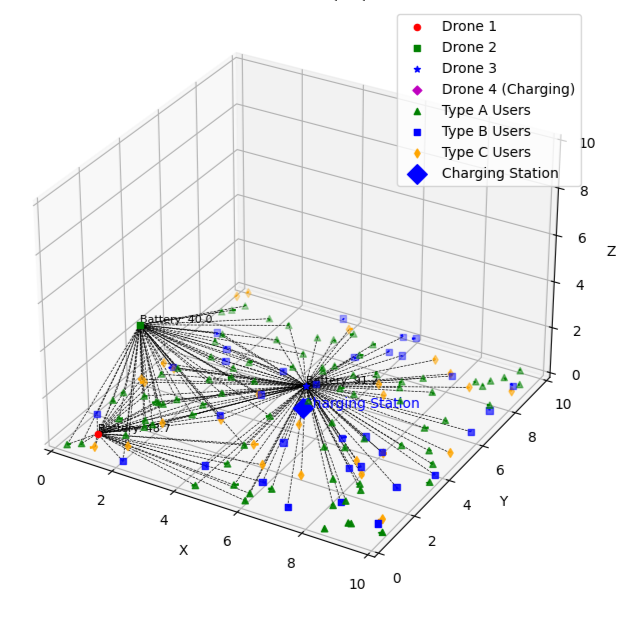}
\caption{\footnotesize MAPPO drone movements in a rural environment: three drones provide coverage while one recharges at the central station. Users are color-coded by priority (high, medium, low), and dashed lines indicate service paths.}
\label{fig:Render}
\end{figure}

\subsubsection{Computational Complexity and Efficiency}

Computational efficiency is critical for real-time MARL deployment in dynamic wireless environments. Table~\ref{tab:comp_efficiency} reports the training time per step and memory footprint for each algorithm in both scenarios.

\begin{table*}[htbp]
\centering
\caption{Computational complexity and memory usage across algorithms}
\begin{tabular}{|c|c|c|c|}
\hline
\rowcolor{lightgray} \textbf{Algorithm} & \textbf{Scenario} & \textbf{Training Time (s/step)} & \textbf{Memory Usage (MB)} \\ \hline
MAPPO   & Urban    & 448.61 & 1588.64 \\ \hline
MAPPO   & Rural    & 393.14 & 1566.54 \\ \hline
MADDPG  & Urban    & 66.96  & 612.31  \\ \hline
MADDPG  & Rural    & 64.57  & 583.88  \\ \hline
MADQN   & Urban    & 25.77  & 360.38  \\ \hline
MADQN   & Rural    & 41.72  & 392.16  \\ \hline
\end{tabular}
\label{tab:comp_efficiency}
\end{table*}

\paragraph{Training time per step}

MADQN is the fastest algorithm, requiring only 25.77~s/step (urban) and 41.72~s/step (rural), making it attractive for computationally constrained or latency-sensitive training pipelines. MADDPG requires moderate computation (approximately 65--67~s/step). MAPPO is the most expensive method by a large margin (approximately 400--450~s/step) due to its centralized critic, multiple gradient-update epochs, and larger batch processing.

\paragraph{Memory usage}

MAPPO also has the highest memory demand (about 1.6~GB), reflecting its richer actor–critic parameterization and larger rollout buffer. MADDPG uses approximately 580--610~MB, while MADQN is the most memory-efficient (360--390~MB). These results indicate that MADQN is best suited for low-resource platforms, while MAPPO is preferable when abundant compute/memory resources are available and high-quality control is required.

\subsubsection{User statistics}

Table~\ref{tab:user_statistics} reports the total number of generated users and the number served at least once in both scenarios.

\begin{table}[htbp]
\centering
\caption{User statistics across scenarios}
\resizebox{\linewidth}{!}{%
\begin{tabular}{|l|c|c|}
\hline
\rowcolor{lightgray} \textbf{User Type / Scenario} & \textbf{Total Generated} & \textbf{Served at Least Once} \\ \hline
Type A (Scenario 1) & 2616 & 2205 \\ \hline
Type B (Scenario 1) & 1504 & 1313 \\ \hline
Type C (Scenario 1) & 532  & 449  \\ \hline
Type A (Scenario 2) & 2251 & 1140 \\ \hline
Type B (Scenario 2) & 1510 & 756  \\ \hline
Type C (Scenario 2) & 524  & 267  \\ \hline
\end{tabular}
}
\label{tab:user_statistics}
\end{table}

Across both environments, the MARL framework successfully serves a substantial fraction of users, with higher service rates for type-$A$ users as intended by the slicing and prioritization logic. These statistics confirm that the learned policies effectively balance user density, signal quality, and energy constraints to prioritize high-demand users while still accommodating types~$B$ and $C$.

%%%%%%%%%%%%%%%%%%%%%%%%%%%%%%%%%%%%%%%%%%%%%%%%%%%%%%%%%%
\section{Conclusion}\label{Sec:conclusion}
%%%%%%%%%%%%%%%%%%%%%%%%%%%%%%%%%%%%%%%%%%%%%%%%%%%%%%%%%%

This work presented a MARL-based framework for UAV-assisted 5G network slicing, integrating MAPPO, MADDPG, and MADQN within a CTDE architecture. By jointly capturing user mobility, TR~38.901-inspired air-to-ground propagation, interference, and UAV energy constraints, the framework enables autonomous and QoS-aware multi-UAV coordination in both urban and rural deployments.
The comparative results highlight a key insight: \textit{no single MARL algorithm is universally optimal}. MAPPO delivers the strongest overall QoS--energy tradeoff, particularly in dense or interference-limited urban environments where coordinated continuous control is critical. MADDPG provides smooth control and competitive SINR in open rural regions but exhibits higher variability and energy usage. MADQN, although computationally efficient, is limited by its discretized action space and achieves weaker QoS performance.
These findings underscore that algorithm choice must be matched to scenario characteristics---including user density, mobility, interference levels, and energy constraints---rather than assuming a one-size-fits-all DRL solution. Future work will investigate hybrid and hierarchical MARL schemes, improved safety and constraint handling, and real-time deployment considerations such as hardware limits, stochastic traffic loads, and larger UAV swarms with onboard sensing. Such extensions will advance scalable, energy-efficient, and QoS-aware UAV communication systems for emerging 6G networks.

\section*{Acknowledgment}
This work was supported by CHIST-ERA under Grant SAMBAS CHISTERA-20-SICT-003 funded by FWO, ANR, NKFIH, and UKRI. This work was also supported by the CNRS through the MITI interdisciplinary programs.
\vspace{11pt}

\bibliographystyle{IEEEtran}
\bibliography{bibliography}

\end{document}